\documentclass[10pt,aps,pra,twocolumn,letterpaper,superscriptaddress,footinbib,longbibliography,floatfix]{revtex4-1}
\usepackage{times,graphicx,amsmath,amssymb,array,sidecap}
\usepackage[hyperfigures,colorlinks=true,allcolors=blue]{hyperref}
\usepackage{ytableau}
\usepackage[utf8]{inputenc}
\usepackage{color}

\newcommand{\NN}{\mathbb{N}}

\newcommand{\CC}{\mathbb{C}}
\newcommand{\SN}{\text{S}_{\!N}}
\newcommand{\UN}{\text{U}(\!N\!\!-\!\!1\!)}

\newcommand{\vc}[1]{\mathbf{#1}}
\newcommand{\abs}[1]{\left|#1\right|}

\newcommand{\ket}[1]{\left|\, #1 \, \right\rangle}

\newcommand{\nspc}{\hspace*{-0.1em}}
\newcommand{\dejm}{ \delta\nspc E_{\text{J}m}}
\newcommand{\dhj}{\delta\nspc H_\text{J}}
\newcommand{\dhu}{\delta\nspc H_U}
\newcommand{\dhc}{\delta\nspc H_\text{C}}
\newcommand{\dhx}{\delta\nspc H_\bullet}
\newcommand{\lag}{\mathcal{L}}

\newcommand{\ejA}{E_{\text{J}}^{\text{a}}}
\newcommand{\ejm}{E_{{\text{J}},m}^{\text{a}}}
\newcommand{\ejbs}{{E_{\text{J}}^{\text{b}}}}

\newcommand{\ecsi}{E_{\text{C}}^{\text{si}}}
\newcommand{\ecA}{{E_{\text{C}}^{\text{a}}}}

\newcommand{\ecbs}{E_{\text{C}}^{\text{b}}}

\newcommand{\ecga}{E_{\text{g}}^{\text{a}}}
\newcommand{\ecgb}{E_{\text{g}}^{\text{b}}}

\newcommand{\tm}{\theta_m}
\newcommand{\tmd}{\dot{\theta}_m}

\newcommand{\phx}{\varphi_\text{ext}}
\newcommand{\phxb}{\bar{\varphi}_\text{ext}}

\newcommand{\closedots}{\cdot \! \! \cdot \! \! \cdot}
\newcommand{\TotalEx}{\textsc{t}}
\newcommand{\st}{ \Theta_\vc{t}}
\newcommand{\sst }{\Sigma^\mu_{\vc{t}}}
\newcommand{\sstl}{\Sigma^\lambda_{\vc{t}}}
\newcommand{\stref}{\Theta^\mu_\text{st}}

\newcommand{\bfr}[1]{\textcolor{red}{\textbf{ #1 } } }

\newcommand{\be}{\begin{equation}}
\newcommand{\ee}{\end{equation}}

\DeclareMathOperator{\sgn}{sgn}

\DeclareMathOperator{\Span}{span}

\begin{document}
\title{Symmetries and collective excitations in large superconducting circuits}
\author{David G.\ Ferguson}
\affiliation{Department of Physics \& Astronomy, Northwestern University, Evanston, IL 60208, USA}
\author{A.\ A.\ Houck}
\affiliation{Department of Electrical Engineering, Princeton University, Princeton, NJ 08544, USA}
\author{Jens Koch}
\affiliation{Department of Physics \& Astronomy, Northwestern University, Evanston, IL 60208, USA}
\date{\today}

\begin{abstract}
In this work we present theoretical tools suitable for quantitative modeling of large superconducting circuits including one-dimensional Josephson junction arrays. The large number of low-energy degrees of freedom, and the peculiar interactions between them induced by flux quantization, present a considerable challenge to the detailed modeling of such circuits. For the concrete example of the fluxonium device we show how to address this challenge. Starting from the complete degrees of freedom of the circuit, we employ the relevant collective modes and circuit symmetries to obtain a systematic approximation scheme. Important circuit symmetries include approximate invariance under the symmetric group and lead to considerable simplifications of the theory.  Selection rules restrict the possible coupling among different collective modes and help explain the remarkable accuracy of previous simplified models. Using this strategy, we obtain new predictions for the energy spectrum of the fluxonium device which can be tested with current experimental technology. 
\end{abstract}
\maketitle

In the search for a viable architecture for solid-state quantum information processing, superconducting circuits have been the focus of immense interest   \cite{Makhlin2001,Devoret2004a,Schoelkopf2008,Clarke2008}. While research efforts have led to the remarkable improvement of coherence times by nearly 5 orders of magnitudes  \cite{Paik2011b} relative to those in the pioneering experiments a decade ago \cite{Bouchiat1998,Nakamura1999}, superconducting circuits have remained extremely simple -- especially when compared to circuits found in commonplace electronic devices. Whether phase, flux, or charge qubits, most superconducting quantum circuits consist of less than a handful of circuit elements. 

Experiments with the fluxonium device -- a superconducting circuit with more than 40 Josephson junctions -- have shown that a larger number of Josephson junctions, and hence degrees of freedom, is not necessarily penalized by reduced coherence times \cite{Manucharyan2009,Manucharyan2010}. Experimental studies of linear Josephson junction arrays have advanced at a rapid pace  \cite{Chow1998,Haviland2000,Takahide2006,Pop2008,Pop2010,Bell2012,Masluk2012}. However, despite considerable theoretical work \cite{Bradley1984,Choi1993,Hermon1996,Odintsov1996,Glazman1997,Choi1998,Matveev2002,Goswami2006,Homfeld2011,Catelani2011a,Rastelli2012} methods for detailed modeling of larger circuit networks are needed to successfully chart the future territory of quantum coherence in networks of increasing size to, e.g., further explore the possibility of topological protection from decoherence \cite{Ioffe2002,Kitaev2006,Gladchenko2008,Doucot2012}. The description presents a considerable challenge to theory due to the combination of several factors: the non-linearity induced by Josephson junctions, the increased number of low energy degrees of freedom, and the peculiar interactions between them induced by flux quantization. As a key step for mastering these difficulties, we present theory for the fluxonium device which starts from the complete circuit degrees of freedom. We demonstrate that circuit symmetries play a crucial role in the organization of the excitation spectrum and, employing the relevant collective modes and their approximate decoupling \cite{Catelani2011a}, we obtain a systematic approximation scheme.

Non-linearity, interactions and a large number of degrees of freedom are challenges commonly encountered in the study of many-electron atoms. 
Our symmetry based approach resembles methods familiar from atomic and molecular physics where the weak breaking of symmetries leads to the well-known lifting of degeneracies in the fine and hyperfine structure of spectra \cite{Condon35}. For the fluxonium circuit, we demonstrate that approximate symmetry under the unitary group and under permutations of junction variables 
divide the excitation spectrum into nearly degenerate subspaces. For realistic parameters, the careful study of perturbations allows us to refine our description and provide new predictions for the collective excitations of the circuit.

\begin{figure}
\centering
	\includegraphics[width=1.0\columnwidth]{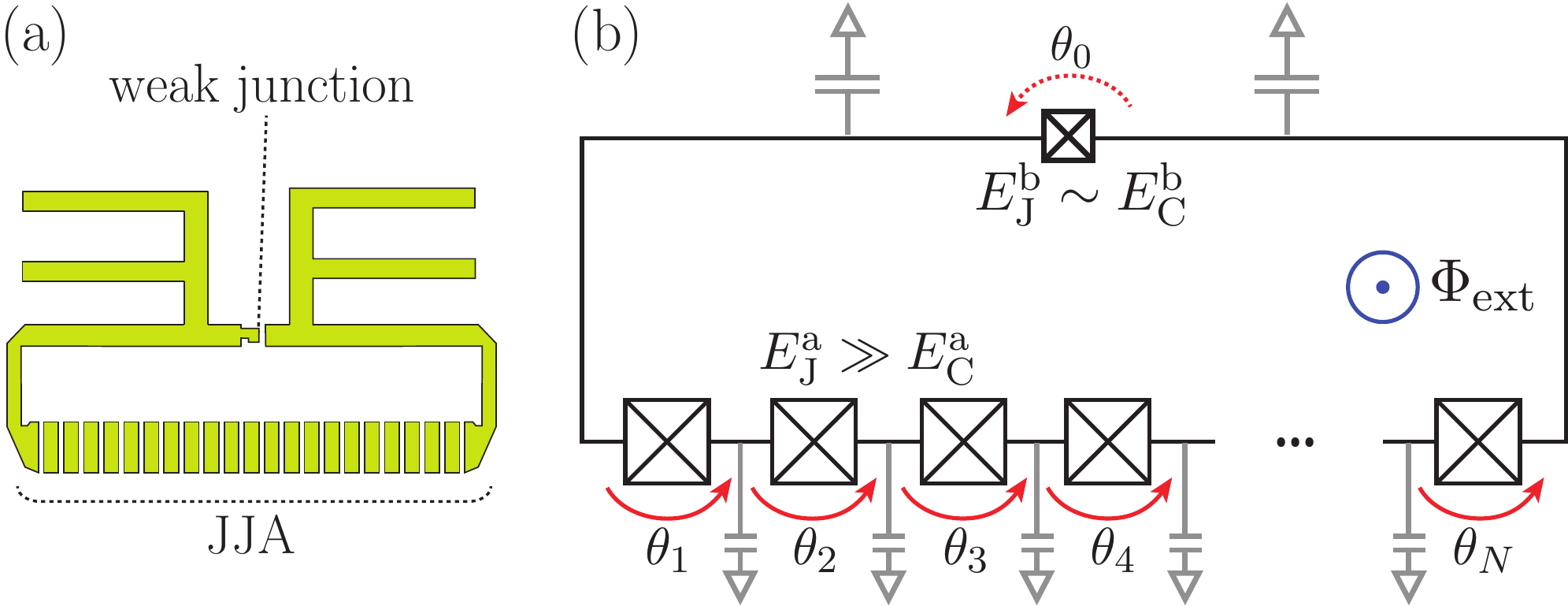}
	\caption{Circuit of the Fluxonium device: Josephson junction array (JJA) of nominally identical junctions with Josephson energy $\ejA$ and charging energy $\ecA$, shunting one weaker junction (Josephson and charging energy $\ejbs$ and $\ecbs$, respectively).}
	\label{fig-circuit}
\end{figure}

\section{Fluxonium circuit}\vspace*{-0.4cm}\noindent
The fluxonium device \cite{Manucharyan2009} (Fig.\ \ref{fig-circuit}) consists of a Josephson junction array with a large number $N\!\gg\!1$ of nominally identical tunneling junctions. One additional  smaller junction (the ``black sheep'') shunts the array. The superconducting loop formed this way can be biased with an external magnetic flux $\Phi_\text{ext}$, making the energy spectrum tunable. As typical of superconducting circuits, the nature of eigenstates and their detailed energy spectrum are governed by the competition between charge transfer across junctions due to Cooper pair tunneling 
and charging effects due to excess electric charge on individual islands. While the former favors charge delocalization and definite phase differences across each junction, the latter promotes charge localization with definite Cooper pair numbers on each superconducting island. The ratio of Josephson energy ($E_\text{J}$) to charging energy ($E_\text{C}$) of the involved junctions quantifies this interplay. We use superscripts ``a" and ``b" in the following to distinguish between array ($\ejA /\ecA \gg 1$) and black-sheep parameters ($\ejbs  \sim \ecbs$).

The energy spectrum and corresponding eigenstates of the superconducting circuit $\mathcal{C}$ are
governed by the stationary Schr\"odinger equation $ H_{\mathcal{C}}\ket\psi=E\ket\psi$, in which the circuit Hamiltonian is obtained from the Lagrangian $\lag=T-U$ by circuit quantization \cite{Devoret1995,Burkard2004}.
For each junction, Josephson tunneling produces a potential energy term $U_m=-E_{\text{J}m} \cos \tm$ where $\tm$ denotes the phase difference across junction $m$. The dominant kinetic energy contributions arise from the charging of junction capacitances, $T_m=\frac{1}{2}C_{\text{J}m} V_{\text{J}m}^2$.
Here, the voltage drop across junction $m$ is linked to $\tmd$ via Josephson's phase evolution equation $\tmd=2\pi V_{\text{J}m}/\Phi_0$, and charging energies are related to capacitances via $E_\text{C} = e^2 /2 C$.

The requirement for the superconducting phase to be single valued (modulo integer multiples of $2\pi$) leads to fluxoid quantization  \cite{London1950,Doll1961,Deaver1961}. It manifests itself as the rigid constraint
$\sum_{m=0}^N\tm+\phx= 2\pi z$ where $z$ is an integer, $\Phi_0=h/2e$ the superconducting flux quantum, and $\varphi_\text{ext}=2\pi\Phi_\text{ext}/\Phi_0$  the phase offset due to external magnetic flux. The constraint reduces the number of independent coordinates by one and induces coupling among the remaining juunction phases.
To incorporate the constraint while maintaining symmetry among array junctions, we eliminate the black-sheep variable $\theta_0$ and obtain the Lagrangian
\begin{align}\nonumber
\lag =& \frac{\hbar^2}{16\ecA} \sum_m  \! \tmd^2+  \frac{\hbar^2}{16\ecbs}\bigg[ \sum_m  \! \tmd\bigg]^2 + \frac{1}{2} \sum_{m n} {\cal G}_{m n} \dot{\theta}_m \dot{\theta}_n \\
&-\sum_m E_{\text{J}m}^\text{a} \cos\tm -\ejbs\cos\bigg(\sum_m \theta_m+\phx\bigg),
\label{lag0}
\end{align}
where, as a convention, sums over Latin indices always run over the range $1,\ldots,N$. The capacitive term involving the matrix ${\cal G}$ describes the effects from capacitances between superconducting islands and ground (see Appendix~\ref{GroundCapAppendix}). 

To illustrate the content of equation \eqref{lag0}, it is instructive to note that $\lag$ describes a single fictitious particle inside a periodic potential, albeit in $N$-dimensional space with $N\gg1$. Alternatively, it can be interpreted as a description of $N$ distinguishable particles, each moving in a 1d periodic potential but subject to a peculiar interaction of collective type induced by flux quantization. 

The central idea of our approach in the following is to harness the large amount of symmetry present in the dominant terms of equation \eqref{lag0} \cite{Note1}. In particular, if ground capacitances are negligible and if all array junctions possess the same charging energy $\ecA$ and Josephson energy $\ejm=\ejA$ then $\lag$ is manifestly $\SN$ symmetric. In other words, any permutation $\sigma\in \SN$ of the array variables, such as
\[
\sigma_{12}[(\theta_1,\theta_2,\theta_3,\ldots,\theta_N)]=(\theta_2,\theta_1,\theta_3,\ldots,\theta_N),
\]
 leaves the Lagrangian invariant for any value of the external flux. We will refer to this idealization as the Symmetric Fluxonium Model (SFM). 

In non-relativistic quantum mechanics, such discrete symmetries generally lead to degeneracies which are governed by the irreducible representations of the symmetry group.  The simplest irreducible representations of the symmetric group $\SN$ are the trivial and alternating representations familiar from particle and many-body physics. In those contexts, they dictate the symmetry of wavefunctions for indistinguishable bosons and fermions. In the case of superconducting circuits,   degrees of freedom referring to different junctions generally remain distinguishable, and the full plethora of irreducible representations of $\SN$ is realized. In this sense, the SFM constitutes an intriguing example of a many-body system with degenerate eigenstates that obey novel permutation symmetries beyond those of bosons and fermions.

\section{$\SN$ symmetric fluxonium model}\vspace*{-0.4cm}\noindent
From the circuit Lagrangian \eqref{lag0} we now extract the relevant collective modes \cite{Catelani2011a} governing the low-energy physics, 
and discuss their connection with effective models employed in previous work \cite{Koch2009,Manucharyan2009}.  
A key ingredient in the construction of the low-energy modes is the observation that array junctions in fluxonium are dominated by Josephson tunneling, $\ecA/\ejA \ll 1$, while the black-sheep parameters $\ecbs\sim\ejbs$ are both roughly of the same order as the array charging energy. 

For large arrays with junction number $N\gg \ejA/\ejbs$, the potential energy $U(\vec{\theta})$ exhibits deep minima at positions where all array coordinates have the identical value 
\be\label{tmseq}
\tm\simeq -\frac{\phxb - 2\pi z}{N+\ejA/\ejbs}
\ee
 with integer $z$ satisfying $\abs{z}\ll N$, and $\phxb=\phx\bmod 2\pi$. 
The minima of $U(\vec{\theta})$ are surrounded by large energy barriers of height $\ge2\ejA$, except along the special direction defined by a simultaneous and equal change in all variables, i.e.\  $\tm = \phi/N$ for all array variables. Such collective dynamics is associated with  the black-sheep variable 
and has a  barrier height of only $2 \ejbs$.
In the quantum regime,  fluctuations will occur primarily along this direction and motivate the use of $\phi$ as an essential collective variable.

\begin{figure}
    \includegraphics[width=0.9\columnwidth]{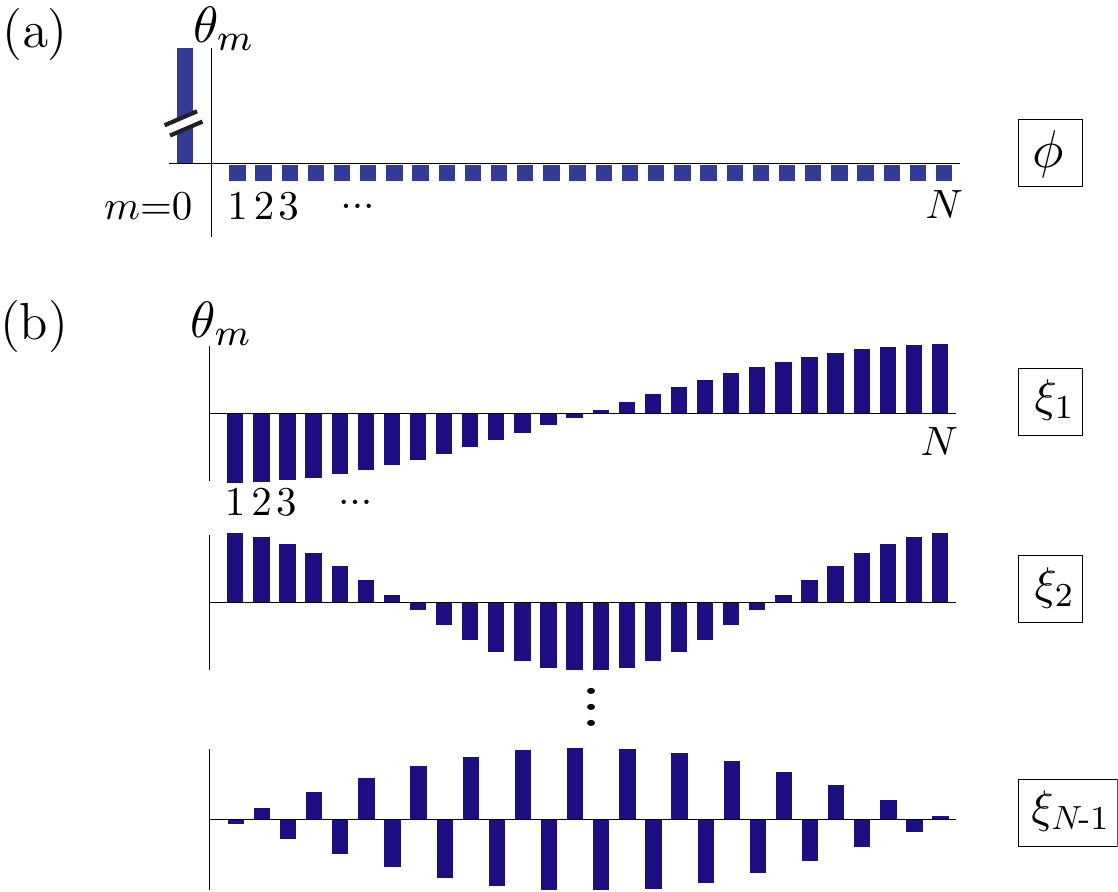}
  \caption{Normal modes for the $\SN$ symmetric fluxonium model. Plots show the array variable amplitudes $\tm$ for each normal mode.  (a) Superinductance mode [$\phi$], for which all array junction amplitudes $\theta_m$ are identical. (b) Difference modes [$\xi_\mu$], for  all of which the amplitude sum exactly vanishes.
  \label{table1}}
\end{figure}

For the Symmetric Fluxonium Model, this collective motion of all array variables forms a normal mode for harmonic oscillations around the global minimum which, for zero magnetic flux, is located at $\phi=0$. Anticipating the role of this mode we refer to it as the \emph{superinductance mode}. Further analysis shows that the remaining $N\!-\!1$ normal modes are degenerate and, so as to remain orthogonal to the superinductance mode, their amplitudes sum to zero mode by mode. We therefore call them \emph{difference modes} and introduce $\xi_\mu$ ($\mu\!=\!1,\dotsc,N\!-\!1$) as  their amplitude variables Fig.~\ref{table1}. The transformation to the new set of variables $\{\phi,\xi_1,\dotsc,\xi_{N-1}\}$ is facilitated by
\be
\label{difference_def}
\tm = \phi/N + \sum_\mu W_{\mu m} \xi_\mu,
\ee 
and inversely, $\phi\!=\!\sum_m \theta_m$ and $\xi_\mu\!=\!\sum_m W_{\mu m} \tm$. Unless otherwise specified, sums over Greek-alphabet indices run over the difference modes $\mu = 1 , \ldots , N-1$.  
The $(N\!-\!1)\!\times\!N$ matrix $W$ is semi-orthogonal and its components sum up to zero in each row, i.e.,
$\sum_m W_{\mu m} W_{\nu m} = \delta_{\mu \nu}$ and $\sum_m W_{\mu m} = 0$.
Our choice 
\be
W_{\mu m} = \sqrt{2/N} \cos \left[\pi \mu (m-\tfrac{1}{2})/N \right]
\ee
differs from the choice in Ref.\ \cite{Catelani2011a}. It proves particularly convenient for the subsequent discussion of corrections from ground capacitances [see Eqs.~\eqref{H_ground_capacitance} and \eqref{Ground_shifts}] which break $\SN$ symmetry. 
After this variable transformation, circuit quantization yields the Symmetric Fluxonium Hamiltonian
\begin{align}\nonumber
H_\text{SFM}=&-4{\ecsi} \partial_\phi^2-4\ecA\sum_{\mu}\partial_{\xi_\mu}^2-\ejbs\cos(\phi + \phx)\\
&-\ejA\sum_{m}\cos\left[\textstyle\phi/N+\sum_\mu W_{\mu m }\xi_\mu\right],
\label{Ham}
\end{align}
where $\ecsi = \ecbs/[1+\ecbs/(\ecA N)]$ is equal to the black-sheep charging energy up to a small $1/N$-correction.
 
The structure of $H_\text{SFM}$ illustrates the utility of the collective-mode description: coupling between different modes is limited to potential energy terms, and the ``effective masses'' are identical for all difference mode amplitudes.
Further, at the relevant potential minima all difference mode amplitudes vanish,  $\xi_\mu\!\!=\!\!0$, and the arguments of the array cosines [last line of equation \eqref{Ham}] are of order $1/N$. Hence, a Taylor expansion for small arguments can be expected to capture the essential low-energy physics.
Keeping terms up to second order in this expansion one obtains
\begin{align*}
H_\text{0} 
= &-4 \ecsi \partial_\phi^2-\ejbs\cos(\phi + \phx) +
\frac{E_L}{2}\phi^2
+ \sum_\mu  \Omega a^\dagger_\mu a^{\phantom{\dagger}}_\mu,
\end{align*}
where $a_\mu^\dagger\!\!=\!\!(\xi_\mu/\Delta_\xi \!-\! \Delta_\xi \partial_{\xi_\mu})/\sqrt{2}$ is the ladder operator creating an excitation in the $\mu$-th difference mode,  $\Delta_\xi\! =\!(8 \ecA /\ejA )^{1/4}$ is the oscillator length,  and $\Omega\!=\!\sqrt{8 \ecA \ejA}$ the array junction plasma frequency. Eigenstates of $H_0$ take the form of direct-product states $ \ket{l}_\text{s}\otimes\ket{\vec{s}}_\text{d}$. In this expression,
$l=0,1,\ldots$ enumerates the superinductance eigenstates (variable $\phi$), and the components of the $(N-1)$ dimensional vector  $\vec{s}$ denote the occupation numbers of the difference modes, i.e.,
\be
 \ket{\vec{s}}_\text{d}=\prod_{\mu=1}^{N-1} \frac{(a_\mu^\dag)^{s_\mu}}{\sqrt{s_\mu !}}|\vec{0}\rangle_\text{d}.
 \ee

The first three terms in the expression for $H_0$ reproduce the superinductance model that was successfully used
in Refs.\ \cite{Manucharyan2009} and  \cite{Koch2009}. It describes the superinductance mode as the coupled system of the black-sheep junction with capacitive energy $\ecsi$ and a large superinductance \cite{Note2} $L_s\!=\!N(\Phi_0/2\pi)^2/\ejA$ with correspondingly small inductive energy $E_L\!=\!\ejA/N$. As the second crucial insight from $H_0$ we note that, within the harmonic approximation, the symmetry of the circuit has been extended to include arbitrary unitary transformations of the $N\!-\!1$ degenerate difference modes. As the superinductance mode is a scalar under the action of the group $\UN$ it  \emph{completely decouples} from the difference modes in the harmonic limit. 
This decoupling explains, in part, the success of the superinductance model in matching experimental spectra in spite of the presence of the large number of additional degrees of freedom. 
The concept of symmetry-induced decoupling carries over to more complicated circuits that include linear arrays of Josephson junctions.

\begin{table*}
\caption{Summary of principal effects of the pertubations $\dhu$, $\dhc$ and $\dhj$ organized by the type of coupling. The three types of coupling are $\text{s}$: coupling amoung superinductance states,   $\text{d}$: coupling amoung difference mode states, and  $\text{sd}$: coupling between the two subsystems.\label{tab:pert-types}}
\centering
\begin{ruledtabular}
\begin{tabular}{ c | p{4.5cm}  p{4.5cm}  p{4.5cm} }
\begin{minipage}{2cm}\bfseries
perturbation\\
origin $\rightarrow$\\ \noindent
\& type $\downarrow$  
\end{minipage}                &\bf  anharmonicity ($\dhu$)  & \bf  capacitance to ground ($\dhc$) &   \bf Josephson energy disorder  ($\dhj$) \\[3mm] \hline 
\\
\textbf{s} & Renormalize $E_L$ [Eq.~\eqref{renormalized_parameters}] & Renormalize $\ecsi$ [Eq.~\eqref{renormalized_parameters}]  & $\qquad\qquad\varnothing$ \\[3mm] 
\textbf{d} & Reduces symmetry from $\UN$ to $\SN$ and splits subspaces into irreducible components $\mathcal{V}_{\textsc{t}} \rightarrow \mathcal{V}_{\textsc{t} (\lambda)} $ [Fig.~\ref{fig-difference_mode_spliting}b]. &  Reduces symmetry from $\SN$ to $PT $ and generates largest energy shift for difference modes with small $\mu$ [Eq.~\eqref{Ground_shifts}]. &  Removes all symmetries and broadens energy distribution of difference modes [Fig.~\ref{fig-spectrum}b].\\[5mm] 
\textbf{sd} &  Symmetry enforced decoupling of subspaces that are inequivalent with respect to $\SN$ symmetry [Fig.~\ref{fig-spectrum}a]. & Creates coupling between superinductor and \emph{even} difference modes [Eq.~\eqref{hcsd} ]. & Creates coupling between superinductor and \emph{all} difference modes. [Eq.~\eqref{hjsd}].
\end{tabular}
\end{ruledtabular}
\end{table*}

\section{Weak $\SN$ symmetry breaking}\vspace*{-0.4cm}\noindent
The Symmetric Fluxonium Model $H_\text{SFM}$ and its approximation $H_0$ both obey $\SN$ symmetry. The symmetries of $H_0$ are enlarged by the harmonic approximation and include an additional  $\UN$ symmetry in the difference-mode subspace: any transformation $a_\mu \to \sum_{\nu} U_{\mu\nu}a_\nu$ with unitary $U$ leaves $H_0$ invariant.
 To go beyond the superinductance model and predict corrections arising from the weak interaction between the superinductance mode and the difference modes, we next consider mechanisms leading to symmetry breaking.

As summarized in Table \ref{tab:pert-types}, we focus on the following three mechanisms which are likely the dominant ones in present experimental samples: anharmonicities of the potential energy neglected in the above expansion ($\dhu$),  disorder in the Josephson energies of individual array junctions ($\dhj$), and  additional stray capacitances of each superconducting island to ground ($\dhc$). We first derive the Hamiltonian expressions for each of these corrections, and discuss their effects on the energy spectrum and eigenstates subsequently.

We start with $\dhu$,  the corrections from anharmonicities exhibited by the periodic potential but neglected in the harmonic approximation employed in  $H_{\text{0}}$. Considering higher-order terms in the Taylor expansion of  $H_\text{SFM} -H_{\text{0}}$, we find that the leading anharmonic corrections are given by
\begin{equation}
    \dhu  = - \frac{\ejA}{4!}\sum_{m}\left(\textstyle \phi/N+\sum_\mu W_{\mu m} \xi_\mu\right)^4.
\end{equation}
It is easy to verify that $\dhu$ breaks the $\UN$ symmetry but preserves the permutation symmetry under $\SN$.

\begin{figure}[b]
\centering
	\includegraphics[width=1.0\columnwidth]{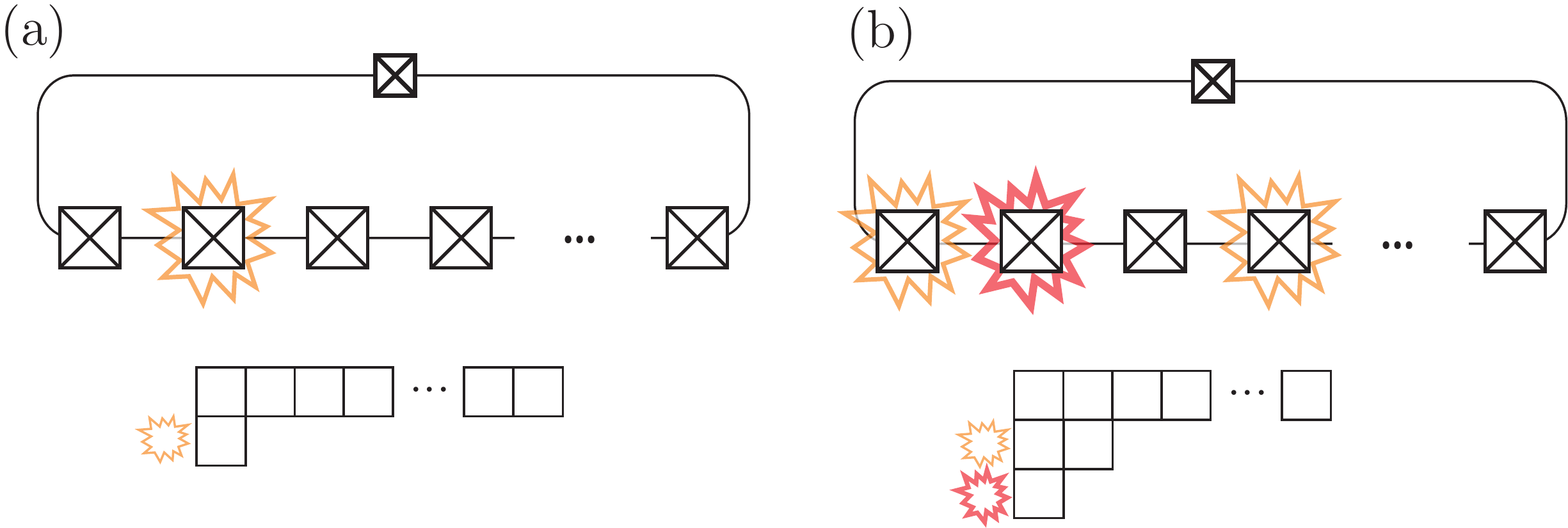}
	\caption[figcaption]{\label{fig-YD-physical} Physical interpretation of Young Diagrams. (a)  Each subspace labeled by a Young diagram of shape $(N$$-$$1,1)$ is spanned by $N$ states in which one array junction is excited relative to the other $N$$-$$1$ array junctions. These states are linearly dependent and can be decomposed into difference mode excitations. For example, the irreducible $(N$$-$$1,1)$ subspace $\mathcal{V}_1$ has a basis comprised of the $N$$-$$1$ difference mode excitations $a_\mu^\dagger {| \vec{0} \rangle}_\text{d}$ which are collective excitations distributed across multiple junctions (see Fig.~\ref{table1}).
(b) Subspaces with higher junction excitations are associated with Young diagrams with additional rows.}
\end{figure}	

To derive an expression for $\dhj$, we capture disorder in the Josephson energies of the array by defining $E_{\text{J}m} = \ejA + \dejm$. Such disorder is expected to be caused by slight variations in junction size and thickness, and may also be affected by junction aging. In the absence of experimental statistics for fluxonium junction parameters, we choose random $\dejm$ from a Gaussian distribution of width $\delta\!E_{\text{J}}=150 \text{MHz}$ and, without loss of generality, impose $\sum_m \dejm = 0$. The disorder modifies the potential energy of the Hamiltonian, and by Taylor expanding
we obtain
\begin{equation}
\label{junction_disorder}
    \dhj 
    = \frac{1}{2}
    \sum_m  \dejm 
    \left( \textstyle
    \phi/N 
    + 
   \sum_\mu  W_{\mu m} \xi_\mu 
    \right)^2.
\end{equation}

Disorder in individual array junction parameters generally leads to weak breaking of both $\UN$ and $\SN$ symmetry.
\begin{figure*}
\centering
\includegraphics[width=0.92\textwidth]{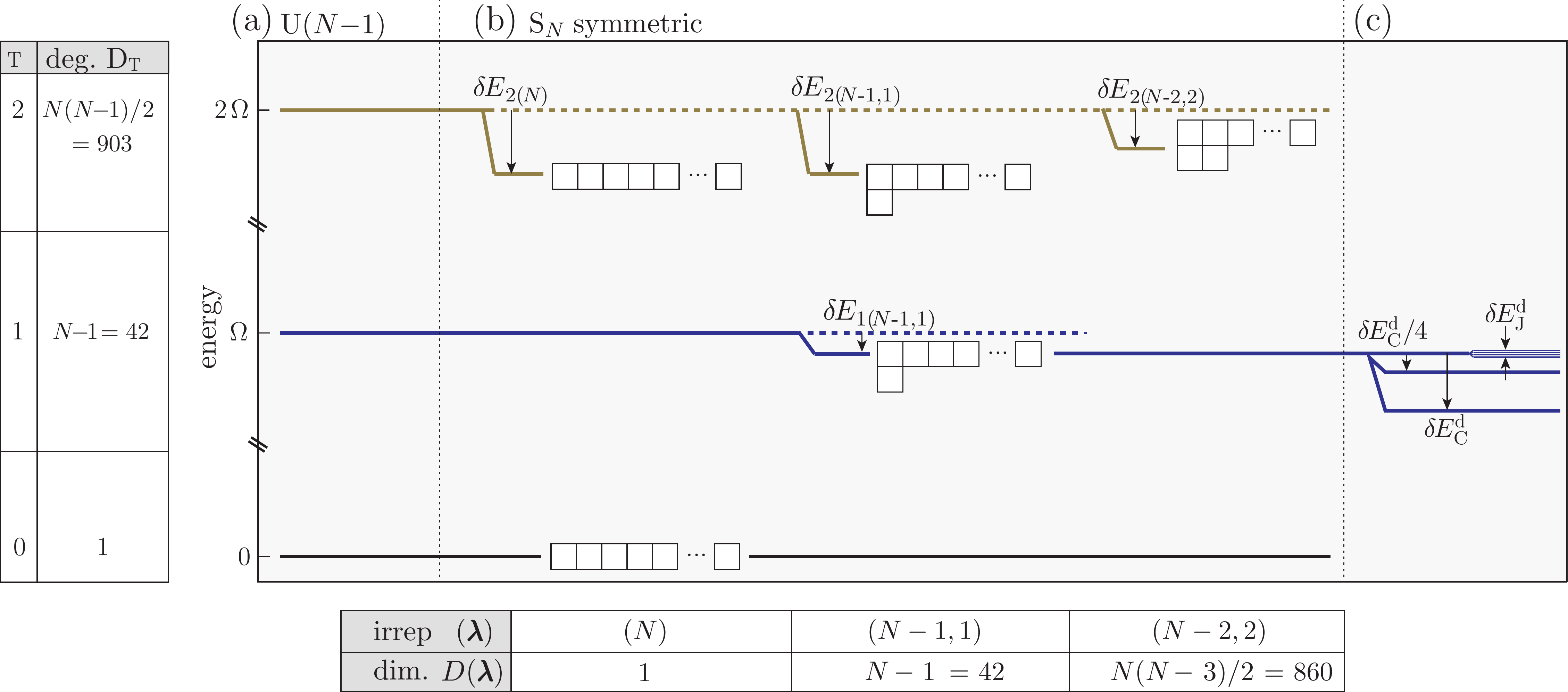}
	\caption{Difference mode spectrum for total excitation numbers $\textsc{t}=0,1,2$. (a)~Within the harmonic approximation, $\SN$ and $\UN$ symmetry hold and produce  degenerate subspaces of dimension $\text{D}_\textsc{t}$. (b)~Anharmonicity $\dhu$ breaks $\UN$ but leaves $\SN$ symmetry  intact. The irreducible representations of $\SN$, labeled by partitions $(\lambda)$ or Young diagrams, give rise to $D(\lambda)$-dimensional degenerate subspaces. (c) Corrections from disorder in array junction $E_\text{J}$s and ground capacitances, $\dhc\! +\!\dhj$ break $\SN$ symmetry. The degeneracy lifting is  shown for the $\textsc{t}\!=\!1$  subspace.\label{fig-difference_mode_spliting}}
\end{figure*}

To capture corrections from stray capacitances of the superconducting islands to ground, we include the terms due to ground capacitances shown in equation \eqref{lag0}. 
Ground capacitances contribute kinetic energy terms which are easily expressed as $T_j = \frac{1}{2} (\Phi_0/2 \pi )^2 C_{\text{g}j} \dot{\varphi}_j^2 $ when using node variables $\varphi_j$ for each superconducting island. Assuming overall charge neutrality of the circuit, we can recast these additional contributions in terms of the junction variables $\tmd$.
Accounting for the ground capacitances of the two large superconducting islands surrounding the black sheep and those of the remaining small islands by  $C_{\text{g}}^{\text{b}},\,C_{\text{g}}^{\text{a}}\ll C_\text{J}^\text{a},\,C_\text{J}^\text{b}$ (Fig.~\ref{fig-circuit}), the perturbation can be expressed as
\begin{equation}
\label{H_ground_capacitance}
    \dhc \simeq  4 \sum_{\mu , \nu = 0}^{N-1}(M^{-1}GM^{-1})_{\mu \nu} \partial_{\xi_\mu}\partial_{ \xi_\nu},
\end{equation}
where $\partial_{\xi_0}\! =\! \partial_{\phi}$ and $\partial_{\xi_\mu} \!=\! (a_\mu - a_\mu^\dagger)/(\sqrt{2} \Delta_\xi)$ for $\mu\ge1$. The detailed derivation of equation \eqref{H_ground_capacitance} and analytical expressions for the entries of the matrices $M$ and $G$ are provided in Appendix~\ref{GroundCapAppendix}. 

After removing irrelevant global energy shifts, the effects of the perturbations $\dhu$, $\dhc$ and $\dhj$ can be organized into three categories according to their action on the superinductance and difference modes. Perturbations may introduce coupling among superinductance states ($\dhx^\text{s}$),  coupling among difference mode states ($\dhx^\text{d}$), as well as coupling between the two subsystems ($\dhx^\text{sd}$). 

\begin{figure*}
\centering
	\includegraphics[width=0.92\textwidth]{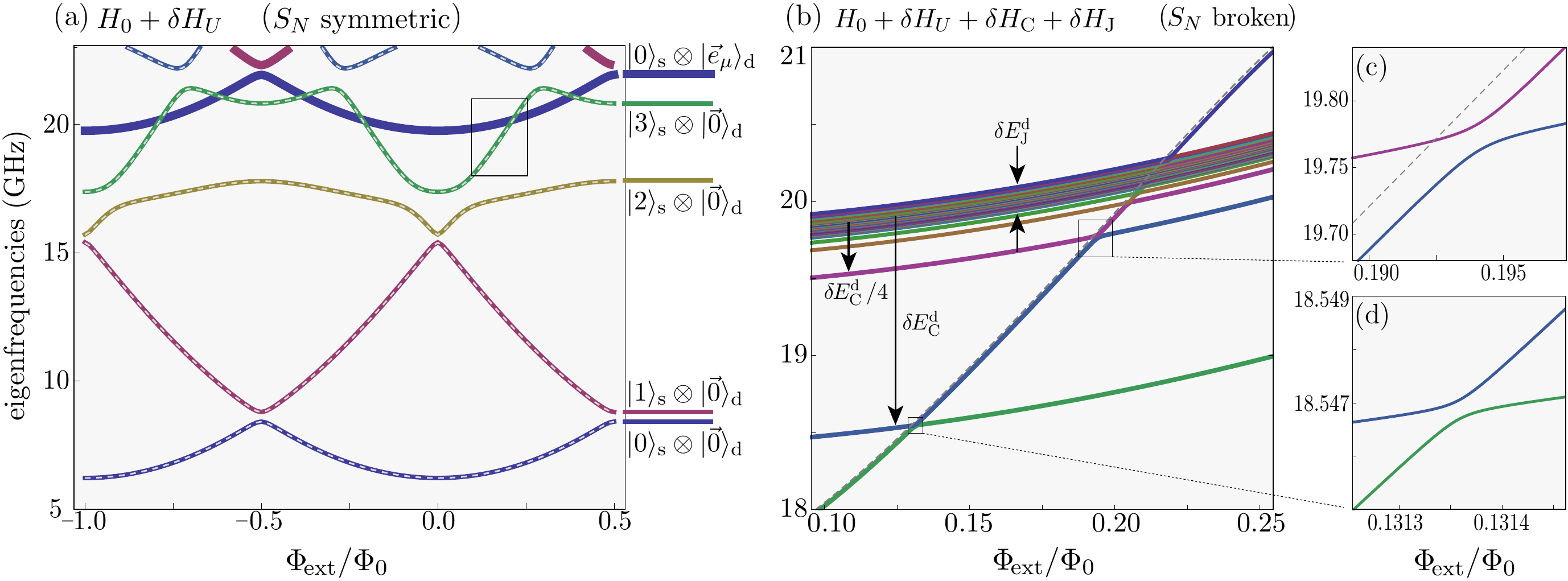}
	\caption[figcaption]{\label{fig-spectrum}Spectrum from numerical diagonalization including effects of (a) anharmonicity,  and (b) anharmonicity, ground capacitances and junction disorder. Dashed curves show the pure superinductance spectrum for renormalized $\ecsi$ and $E_L$ [equation \eqref{renormalized_parameters}]. Thick curves in (a) indicate $N\!-\!1$-fold degenerate levels that remain decoupled from $\textsc{t}=0$ states under $\dhu$.  (b) Corrections from ground capacitances $\dhc$ and junction disorder $\dhj$ break $\SN$ symmetry. In the $\textsc{t}=1$ manifold, ground capacitances split off several levels; smaller shifts are due to junction disorder $\delta\nspc E_{\text{J}\mu}^\text{d}$, here chosen from a Gaussian distribution. The panels in (c), (d)  show magnifications of regions marked in (b). The avoided crossing in (c) is primarily generated by ground capacitances. The even smaller splitting in (d) is purely generated  by array junction disorder. Parameters are chosen consistent with experimental device (Appendix~\ref{ParamValues}).}
\end{figure*}

We first discuss corrections in the $\dhx^\text{s}$ category, which only affect the superinductance mode. The simplest contributions of this type are terms with a structure identical to those already present in $H_0$, which merely renormalize the superinductance model parameters. Both $\dhu$ and $\dhc$ contain corrections of this type and yield renormalized  parameters
\begin{align}\nonumber
\ecsi &\to \frac{1}{ 1/\ecbs+1/N\ecA + G_{00}},\\
E_L &\to \frac{\ejA}{N}\left[ 1-\Delta_\xi^2\frac{N-1}{4N} \right].
\label{renormalized_parameters}
\end{align}
We emphasize that the inclusion of ground capacitances of the large islands alone leave the $\SN$ symmetry unharmed and are completely accounted for in our model by the above renormalization, even in the limit that $C_\text{g}^\text{b}$ is large. The only contribution of type $\dhx^\text{s}$ which goes beyond renormalization is the term $- \ejA \phi^4/(4! N^3)$ generated by $\dhu$. For realistic parameters, we find that this perturbation generates level shifts well below $100 \, \text{MHz}$ (Fig.~\ref{fig-spectrum}a).

\begin{table}
\caption{Energy corrections for difference mode states due to anharmonicities ($\dhu$). $\textsc{t}$ is the total number of excitations in the difference modes and $(\lambda)$ the partition labeling the irreducible subspace.\label{tab:corr}}
\centering
\begin{ruledtabular}
\begin{tabular}{ccc}
$\textsc{t}$ & $(\lambda)$ & $-\delta\nspc E_U^\text{d}[\textsc{t},(\lambda)]/\ecA$\\\hline
$0$ & $(N)$            & 0 \\
$1$ & $(N\!-\!1,1)$ & $1-1/N$ \\
$2$ & $(N)$            &  $3-3/N$\\
$2$ & $(N\!-\!1,1)$ & $3-4/N$ \\
$2$ & $(N\!-\!2,2)$  & $2-2/N$
\end{tabular}
\end{ruledtabular}
\end{table}

We next turn to corrections in the $\dhx^\text{d}$ category to discern how anharmonicity $\dhu$, ground capacitance $\dhc$, and array junction disorder $\dhj$ affect the spectrum of difference mode excitations.
Perturbations from anharmonicity $\dhu$ break the $U(N-1)$ symmetry but leave the $\SN$ symmetry subgroup intact. As a result, degeneracies are lifted only partially and degenerate perturbation theory must be used. Each remaining degenerate subspace is associated with an irreducible representation of the symmetric group. Our construction of the relevant irreducible subspaces works as follows. We start by decomposing the difference-mode Hilbert space into orthogonal subspaces $\mathcal{V}_\textsc{t}$ with fixed excitation number $\textsc{t}=\sum_\mu a_\mu^\dag a_\mu$, i.e.,
$\mathcal{H}_\text{d}=\mathcal{V}_0\oplus\mathcal{V}_1\oplus\cdots$. In general, each $\mathcal{V}_\textsc{t}$ may still be reducible under $S_N$ and should be decomposed further.

In this decomposition, the integer partitions of $N$ serve as labels for the irreducible representations of the symmetric group $\SN$. Here, a partition $(\lambda) = (\lambda_1,\lambda_2,\dotsc,\lambda_F)$ is a sequence of  non-increasing positive integers 
\[ \lambda_1\ge \lambda_2 \ge \cdots \ge \lambda_F>0 \]
 that sum to $N$.
Each partition is conveniently represented by a Young diagram: a collection of $N$ boxes arranged in $F$ left-justified rows with the i$^\text{th}$ row having the length $\lambda_i$. For $N=6$ the partition $(4,2)$ is represented by the Young diagram
\[
(4,2) 
\quad \leftrightarrow \quad 
{\ytableausetup{centertableaux,boxsize=0.9em}
	\begin{ytableau}
		\phantom{0} &  \phantom{0} & \phantom{0} & \phantom{0} \\
		\phantom{0} & \phantom{0}
	\end{ytableau}} \, .
\]
Since the inductive decomposition of $\SN$ \cite{Hamermesh1989} is not very practical for  $N\!\gg\!1$, we decompose the subspaces  ${\mathcal{V}}_{\textsc{t}}$ by using a restricted set of semi-standard Young tableaux  \cite{Sagan}. (All technical details of this procedure are provided in Appendix~\ref{DiffModeDecomp}.) For the low-energy part of the spectrum probed by experiments, we find that the excitation number $\textsc{t}$ and partition $(\lambda)$ are sufficient to specify the relevant irreducible subspaces.  A simplified physical interpretation of the Young diagrams is offered in Fig.~\ref{fig-YD-physical}.

The subspace without any difference mode excitations, $\mathcal{V}_{\textsc{t}=0}$, is spanned by only the ground state ${|\vec{0} \rangle }_\text{d}$. 
It immediately forms an irreducible representation. The state is effectively bosonic and is indexed by the partition $(\lambda) = (N)$. 
The subspace of difference mode states with a single excitation, $\mathcal{V}_{\textsc{t}=1} = \text{span} \{ \, a^\dagger_\mu { | \vec{0} \rangle }_\text{d} \, \, | \, \, \mu=1, 2, \ldots, N-1 \}$
already forms an irreducible $N\!-\!1$-dimensional subspace corresponding to the representation with partition $(\lambda) = (N\!-\!1,1)$  (Fig.\ \ref{fig-difference_mode_spliting}). For $\textsc{t} = 2$ the decomposition is more interesting and results in three irreducible subspaces indexed by $(N)$, $(N\!-\!1,1)$ and $(N\!-\!2,2)$. The  subspace labeled by $(N)$, for example, is comprised of a single $\SN$ invariant state given by 
\begin{equation}
\ket{\textsc{t}(\lambda)}=\ket{2(N)} = \frac{1}{\sqrt{2 (N-1)}}\sum_{\mu} (a^\dagger_\mu)^2 {| \vec{0} \rangle}_\text{d},
\end{equation}
which is independent of the specific choice of $W$ in equation \eqref{difference_def}. By employing perturbation theory for $\dhu^\text{d}$ in each irreducible subspace $\mathcal{V}_{\textsc{t}(\lambda)}$, we obtain the first order energy corrections $\delta\nspc E_U^\text{d}[\textsc{t},(\lambda)]$ in Table \ref{tab:corr}. The resulting level shifts are shown schematically in Fig.~\ref{fig-difference_mode_spliting}b.

We next consider perturbations which break the $S_N$ symmetry and thus lift the degeneracy of difference modes. Both corrections from ground capacitances, $\dhc^\text{d}$,  and junction disorder, $\dhj^\text{d}$, fall in this category.  If the  ground capacitance of the big islands is sufficiently large compared to that of the small islands, $\ecgb / \ecga  \ll 1/N^2$, then $\dhc^\text{d}$ in  equation \eqref{H_ground_capacitance} is approximately diagonal. This decoupling is the primary motivation for our choice of difference modes [Eq.~\eqref{difference_def}].  To leading order, the resulting energy shifts for states in the in the 1-excitation subspace ${\mathcal{V}}_1$ are given by
\begin{equation}
\label{Ground_shifts}
\delta\nspc E_{\text{C},\mu}^\text{d} 
 \approx -\delta\nspc E_{ \text{C} }^\text{d} / \mu^2 \qquad (\mu \ll N),
\end{equation}
where  $\delta\nspc E_{\text{C}}^\text{d}\! = \!4 N^2 \ecA^2  /(\pi^2 \Delta_\xi^2 \ecga )$. Similar effects from ground capacitances have been predicted and observed in Ref. \cite{Masluk2012}.
For higher values of $\mu$, shifts induced by Josephson energy disorder $\dhj^\text{d}$ become dominant in our model. For Gaussian distributed junction parameters $\dejm$,  the first-order energy shifts $\delta\nspc E_{\text{J},\mu}^\text{d}$ also follow a Gaussian distribution with width $	\delta E_{\text{J}}^\text{d} = \delta\nspc E_{\text{J}} \Delta_\xi^2/2$ (Figs.~\ref{fig-difference_mode_spliting}c and~\ref{fig-spectrum}b).

Interesting corrections in the third and final category, $\dhx^\text{sd}$, arise from coupling between superinductance and difference modes. Anharmonicity captured by $\dhu^\text{sd}$ preserves $\SN$ symmetry and hence, by Schur's lemma, cannot couple states belonging to different irreducible representations. More specifically, states of the form $ { | \ell' \rangle }_\text{s} \otimes a^\dagger_\mu { | \vec{0} \rangle }_\text{d}$ cannot couple to states of the form $ {| \ell \rangle}_\text{s} \otimes {| \vec{0} \rangle }_\text{d}$ under $\SN$ symmetry, even when such states are degenerate (Fig.~\ref{fig-spectrum}a). 
This symmetry-enforced lack of coupling between the superinductance mode and the lowest difference-mode excitations constitutes the second central result of our work. It is a crucial ingredient in preserving the respective identity of these collective modes and explains the quantitative accuracy of the superinductance model at low energies. The only difference-mode excitations that \emph{may} couple to states of the form ${|\ell \rangle}_\text{s} \otimes {| \vec{0} \rangle }_\text{d}$ are those that are bosonic, i.e., are indexed by the partition $(N)$. The candidate states with lowest energies are ${ | \ell' \rangle}_\text{s} \otimes {| 2 (N) \rangle }$ but are already well beyond the frequency range probed by spectroscopy in previous fluxonium experiments. 

Ground capacitances, as described by $\dhc^\text{sd}$, break $\SN$ symmetry but preserve $PT$ symmetry: $\varphi_j \rightarrow -\varphi_{N-j}$. $PT$ symmetry is a combination of ``circuit parity" $P$ which mirrors the circuit variables, and time reversal $T$. $PT$ is a symmetry of $H_0$ and $H_{\text{SFM}}$, even for non-zero flux $\varphi_{\text{ext}}$. The superinductance mode and difference modes with even index $\mu$ are {\emph{even}} under $PT$; difference modes with odd index $\mu$ are {\emph{odd}} under $PT$. As a result,  $\dhc^\text{sd}$ can only couple the superinductance mode to every other difference mode. We find that the coupling is largest for small values of $\mu$ and takes the form

\be\label{hcsd}
 \dhc^\text{sd} \approx -\partial_\phi \sum_{\mu=2,4,6\cdots} \delta\nspc E_{\text{C}}^\text{sd} (a_\mu - a^\dagger_\mu)/\mu^2 ,
\ee

where $ \delta\nspc E_{\text{C}}^\text{sd} \!= \!8 \ecA \ecsi N^{3/2} /(\pi^2 \Delta_\xi \ecga )$.
Finally, all symmetries are broken for array junction disorder, and the resulting perturbation is given by

\be\label{hjsd}
 \dhj^\text{sd} \approx  \phi \sum_\mu \delta\nspc E_{\text{J}\mu}^\text{sd} (a_\mu + a^\dagger_\mu), 
\ee

with $\delta\nspc E^\text{sd}_{\text{J},\mu}$ following a Gaussian distribution of width $ \delta\nspc E_{\text{J}} \Delta_\xi / (\sqrt{2} N)$.
As shown in Figs.~\ref{fig-spectrum}c,d, the coupling between  superinductance and the $\mu =1$ difference mode, induced by $\dhj^\text{sd}$ only, is considerably smaller than the coupling to the $\mu = 2$ difference mode which is dominated by $\dhc^\text{sd}$.

\section{Discussion and Summary}\vspace*{-0.4cm}\noindent
The low-energy spectrum of the full fluxonium circuit includes, in addition to the energy levels predicted by the superinductance model a large number of nearly degenerate excitations. We have identified the nature of these collective excitations with the difference modes at energies near the array junction plasma frequency. Degeneracies are expected to be lifted, first by $\UN$ symmetry breaking due to anharmonicity 
and further by $\SN$ symmetry breaking due to array junction disorder and ground capacitances. The important consequences of these corrections include separation of previously degenerate levels into closely spaced multiplets.

Josephson junction arrays provide an interesting example of a quantum system with many identical but \emph{distinguishable} degrees of freedom, resulting in representations of the symmetric group not readily observed in nature with indistinguishable particles.
Invariance under permutations of the junction variables is a generic symmetry expected to be important for any large superconducting circuit containing one or several Josephson junction arrays. The decomposition of the symmetric group $\SN$ into irreducible representations relevant at low energies thus becomes an important tool in circuit analysis. For the example of the fluxonium device, we have shown that such symmetry strongly restricts the possible coupling between the superinductance mode, as observed in the experiment \cite{Manucharyan2009,Manucharyan2010}, and the additional difference modes. Our results explain the remarkable accuracy of the effective superinductance model  as long as the renormalizations of $E_L$ and $\ecsi$ are taken into account, and are consistent with previous fits of experimental data where $E_L$ and $\ecsi$ were used as a fit parameters, producing excellent agreement  \cite{Manucharyan2009,Manucharyan2010}.

The power of symmetry-based approaches in the analysis of future circuits is easily illustrated for the example of the fluxonium device. Specifically, the number of difference-mode states with excitations up to some threshold $\textsc{t}$ grows rapidly as $\frac{(\textsc{t}+N-1)!}{\textsc{t}!(N-1)!}$. The number of states with proper bosonic symmetry, however, is dramatically smaller:  For $N=43$ and $\textsc{t}=5$ there are $10^6$ difference mode states states but merely $6$ of them possess  bosonic symmetry. Harnessing exact and approximate symmetries of Hamiltonians for larger circuits will likely be a crucial ingredient in future research exploring quantum coherence in superconducting circuit networks of increasing complexity.

\begin{acknowledgments}
We thank Leonid Glazman, Michel Devoret and Vladimir Manucharyan for stimulating discussions.  Our research was supported by the NSF under Grants PHY-1055993 (JK, DGF), DMR-0805277 (DGF), and by the David and Lucile Packard Foundation (AAH).
\end{acknowledgments}


\appendix
\noindent
\section{Irreducible representations for difference modes}
\label{DiffModeDecomp}
In this appendix we discuss the decomposition of the difference-mode Hilbert spaces $\mathcal{V}_{\TotalEx}$ into subspaces that transform irreducibly under $\SN$ symmetry. Since some of the mathematical tools employed may not belong to the physicist's ordinary repertoire, we provide definitions along with concrete examples where appropriate.  In terminology and notation, our discussion closely follows the excellent book by Sagan  \cite{Sagan}.

The subspace $\mathcal{V}_{\TotalEx}$ comprises all difference-mode states with total excitation number $\TotalEx$. It is spanned by the orthogonal states
\begin{equation}\label{dmstates}
a^\dagger_{\mu_1}\cdots a^\dagger_{\mu_{\TotalEx}} |\vec{0}\rangle
\end{equation}
where we assume weakly ordered mode indices 
\[ 
\mu_1 \le \cdots \le \mu_{\TotalEx} \in \{1,2,\cdots,N-1\}
\]
to avoid double counting.
The $\SN$ symmetry displayed in the ideal fluxonium circuit pertains to permutations  $\sigma \in \SN$ of the array junction variables $\theta_1,\theta_2,\ldots,\theta_N$. Such permutations also induce linear transformations in the operator space $\CC\{a_1^\dag,a_2^\dag,\ldots,a_{N-1}^\dag\}$ spanned by the difference-mode creation operators. To understand how $a_\mu^\dag$ transforms under permutations, we recall the definition of the creation operators in terms of junction variables:
\be
a_\mu^\dag = \sum_m W_{\mu m} (\theta_m/\Delta_\xi - \Delta_\xi \partial_{\tm})/\sqrt{2}.
\ee
Using the identity $\sum_\mu W_{\mu m} W_{\mu n} =	\delta_{mn} - \frac{1}{N}$, one finds that the difference-mode creation operators transform according to 
\be
\label{a_transform}
\sigma({a}^\dagger_\mu)= \sum_\nu S(\sigma)_{\mu \nu} {a}^\dagger_\nu.
\ee
We remind the reader that, by our convention, sums with Latin (Greek) summation indices always range from $1$ to $N$ ($1$ to $N-1$).
The $(N\!-\!1)\!\times\!(N\!-\!1)$ transformation matrices are given by
\be
S(\sigma)_{\mu \nu} = \sum_{m,n} W_{\mu m} W_{\nu n} D(\sigma)_{mn}. 
\ee
Here, $D(\sigma)_{m n}= \delta_{m,\sigma(n)}$ denotes the $N\!\times\!N$ permutation matrix for the group element $\sigma\in\SN$. (The matrices $D(\sigma)$ form the defining representation of $\SN$.)

By the relation 
$\mathcal{V}_1=\CC\{a_1^\dag,a_2^\dag,\ldots,a_{N-1}^\dag\}|\vec{0}\rangle$, the transformation matrices $S(\sigma)$  in equation \eqref{a_transform} define
an orthogonal $\SN$ representation of degree $N\!-\!1$ in the one-excitation subspace. 
Similarly, for higher excitation numbers $\TotalEx >1$, the group action for products of creation operators,
\be
\sigma \left( {a}^\dagger_{\mu_1} \cdots {a}^\dagger_{\mu_{\TotalEx}} \right)= \sum_{\nu_1, \cdots, \nu_\TotalEx } S(\sigma)_{\mu_1 \nu_1} \cdots S(\sigma)_{\mu_\TotalEx \nu_\TotalEx} {a}^\dagger_{\nu_1} \cdots {a}^\dagger_{\nu_{\TotalEx}},
\ee
determines the representation of $\SN$ in the subspace $\mathcal{V}_\TotalEx$. Given these representations, our central task is to decompose each $\mathcal{V}_\TotalEx$ into its irreducible subspaces. As an aside we note that in the special case of $\mathcal{V}_1$, simple arguments based on group characters can be used to show that $\mathcal{V}_1$ is already irreducible and coincides with the irreducible representation indexed by the partition $(N \! - \! 1,1)$, for which group characters are known to be  $\text{tr}D(\sigma) -1$ for arbitrary $N$ (Ref.\  \cite{Sagan} section 2.12).

The common approach for decomposition of such product representations is inductive and requires successive decompositions for $\text{S}_{1},\,\text{S}_{2},\ldots,\SN$, see, e.g., Ref. \cite{Hamermesh1989}. For large $N$, however, that strategy is not very practical. Following the treatment by Sagan  \cite{Sagan}, we thus employ an alternate approach using a restricted class of semi-standard tableaux. (We explain the meaning of these words in due course). 

 As our first step in constructing the decomposition of each $\mathcal{V}_\TotalEx$, we define the pseudo-creators $b_n^\dag$ for $n=1,2,\ldots, N$  by
\begin{align}\label{pseudob}
b_n^\dag &= \sum_\mu W_{\mu n} a_\mu^\dag \\\nonumber
&=  \left[ \left(\theta_n - \phi/N\right)/\Delta_\xi + \Delta_\xi \left(\partial_{\theta_n} - \partial_\phi \right)\right]/\sqrt{2}.
\end{align}
As one would expect, pseudo-creators $b_n^\dag$ increase the total excitation number in the difference-mode subspace by one. The number of pseudo-creators, however, is $N$ and thus exceeds the number of difference-mode creation operators $a_\mu^\dag$ by one. Indeed, the pseudo-creators obey $\sum_n b_n = 0$ and, hence, are not linearly dependent. They obey the non-standard commutation relation
\be
\label{b_commutation}
[b_m,b^\dagger_n] = \delta_{mn} - 1/N.
\ee
For the price of this anomalous commutator, we obtain operators which transform with elegant simplicity. Specifically, under array variable permutations $\sigma\in\SN$, the $b_n^\dag$ operators simply undergo  permutations:
\begin{equation}\label{pmeq}
\sigma({b}_n^\dag) = \sum_m D(\sigma)_{n m} b_{m}^\dag= b_{\sigma(n)}^\dag.
\end{equation}
This simple transformation law will be crucial for finding the irreducible subspaces of $\mathcal{V}_\TotalEx$.

We next extend the language of difference-mode excitations to pseudo-mode excitations and define the states
\begin{equation}
    |\vc{t} \rangle = \prod_{n=1}^{N} (b_n^\dag)^{t_n} | \vec{0} \rangle\qquad\text{(not normalized)},
\end{equation}
where the vector $\vc{t} = (t_1,t_2,\cdots,t_N)$ specifies the excitation numbers $t_n\in\NN_0$ for each pseudo-mode $b^\dag_n$.
Using the inverse of equation \eqref{pseudob}, $a_\mu^\dag = \sum_n W_{\mu n} b_n^\dag$, it is simple to confirm that 
\be
\mathcal{V}_\textsc{t} = \Span \{ \ket{\vc{t}} \, | \,\textstyle \sum_n t_n=\textsc{t} \},
\ee
i.e., the pseudo-mode excitations span the difference mode subspaces $\mathcal{V}_\textsc{t}$ one by one. For example, the pseudo-mode states $b_n^\dag | \vec{0} \rangle$ span the irreducible subspace $\mathcal{V}_1$ indexed by $(N$$-$$1,1)$ (see Fig.~\ref{fig-YD-physical}).

For a state with given pseudo-mode excitations $\vc{t} = (t_1,t_2,\cdots,t_N)$, equation \eqref{pmeq} implies that $\sigma\in\SN$ simply permutes the pseudo-mode excitation numbers according to $\vc{t}\to\vc{t}'= (t_{\sigma(1)},t_{\sigma(2)},\cdots,t_{\sigma(N)})$.
\begin{quote}
\emph{Example:} The state with $3$ excitations in pseudo-mode $n=1$ and $4$ excitations in pseudo-mode $n=2$ is $\vc{t}=(3,4,0,\ldots,0)$. A permutation may transform it into $\vc{t}'=(4,0,3,0,\ldots,0)$, for example but not into $(2,5,0,\ldots,0)$, even though the latter state still has the same total excitation number $\textsc{t}=7$. 
\end{quote}
For each state $|\vc{t}\rangle$ with  $\textsc{t}=\sum_n t_n$, we define the subspace  spanned by itself and its permuted partner states:
 \be
   \mathcal{V}_{[\vc{t}]} := \Span\left\lbrace 
    | \vc{t}' \rangle=|\sigma\vc{t}\rangle  \, \mid \,  \sigma\in\SN \right\rbrace \subset \mathcal{V}_\textsc{t},
 \ee
As suggested by the notation, $[\vc{t}]$ may be understood as an equivalence class when defining $\vc{t}\sim\vc{t}'\,:\Leftrightarrow\,$there exists a $\sigma\in\SN$ such that $\vc{t}'=\sigma\vc{t}$.
By construction, the $ \mathcal{V}_{[\vc{t}]} $ form $\SN$-invariant subspaces, and their unions cover each  $\mathcal{V}_\TotalEx$
\be
	\mathcal{V}_\TotalEx = \bigcup_{ \left\{ [\vc{t}] \, | \, \TotalEx = \sum_n t_n \right\} } \mathcal{V}_{[\vc{t}]}.
\ee
Note that, due to linear dependence of the pseudo-modes, subspaces for inequivalent excitation classes $[\vc{t}]\cap[\vc{t}']=\emptyset$ may, nonetheless, have a non-zero intersection, $\mathcal{V}_{[\vc{t}]}\cap\mathcal{V}_{[\vc{t}']}\not=\emptyset$. 

We will first discuss the decomposition of $ \mathcal{V}_{[\vc{t}]}$ into irreducible subspaces as if $b_n$ were orthogonal modes.  In step 1, we thus drop the cautionary prefix ``pseudo" temporarily and show  that the basis vectors $\ket{\vc{t}}$ can then be relabeled in such a way to reveal isomorphism between $ \mathcal{V}_{[\vc{t}]}$ and the corresponding permutation module $M^{\lambda_\vc{t}}$. In step 2, we then utilize the important theorem for the decomposition of $M^{\lambda_\vc{t}}$ that identifies semi-standard tableaux as the indexing set for all irreducible subspaces. In both steps, we introduce the necessary terminology and explain the construction. We do not provide proofs of the underlying theorems but refer the interested reader to Sagan's book  \cite{Sagan}, chapter 2. Finally, in step 3 we return to the issue
of linear dependence of $b_n$ modes and show how the usual construction can be modified to account for the linear dependence in a simple fashion.

\begin{figure}
\centering
	\includegraphics[width=0.95\columnwidth]{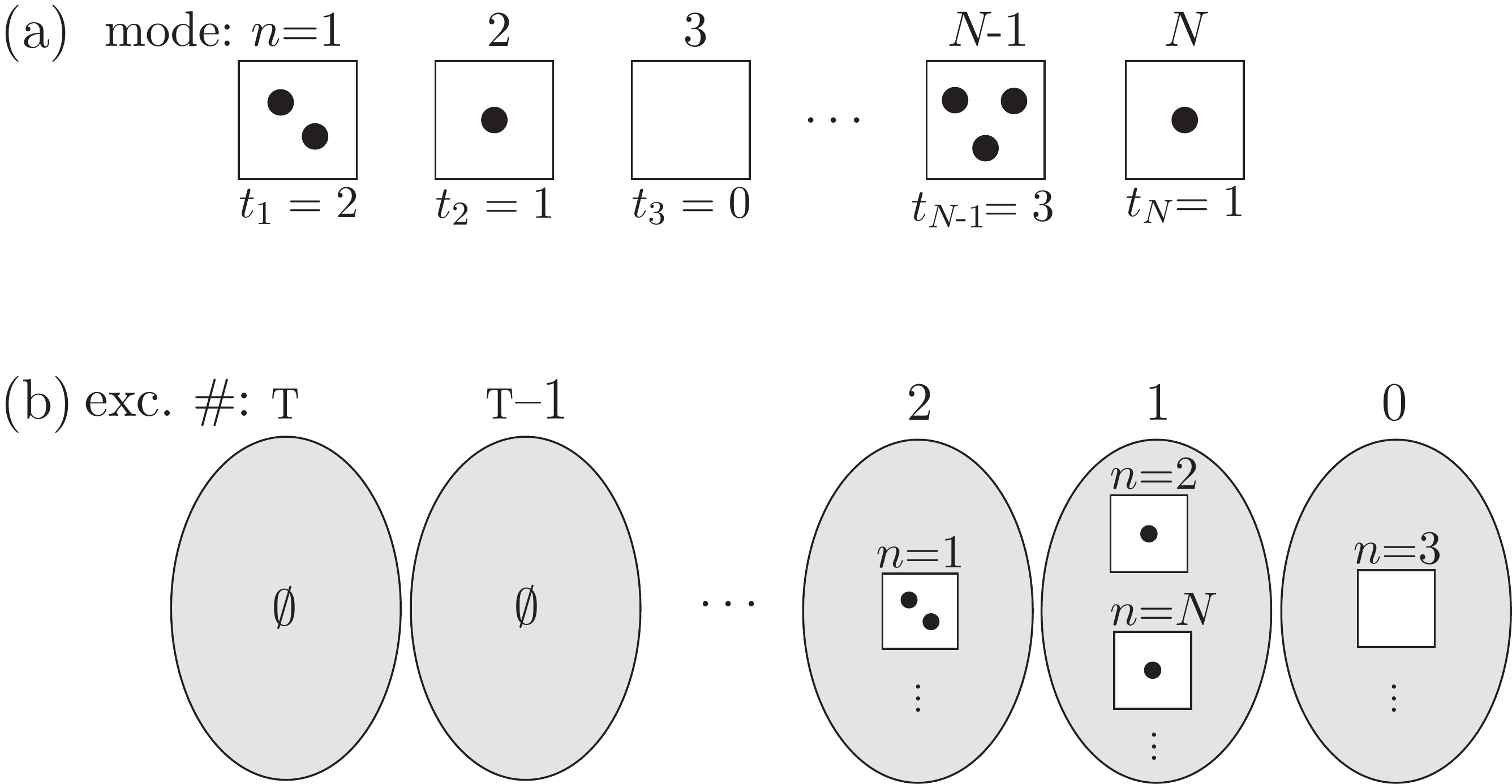}
	\caption{Labeling schemes for difference mode excitations. (a) Label specifies excitation numbers, mode by mode. (b) Label specifies mode indices, excitation number by excitation number.
	\label{fig:labeling}}
\end{figure}

\subsection*{STEP 1: Isomorphism between $ \mathcal{V}_{[\vc{t}]}$ and the permutation module $M^{\lambda_\vc{t}}$}
The excitation numbers $\vc{t}=(t_1,t_2,\ldots,t_N)$ label each state in  $ \mathcal{V}_{[\vc{t}]}$ by specifying \emph{excitation numbers, mode by mode}. An alternative labeling scheme (Fig.\ \ref{fig:labeling}) consists of specifying \emph{mode numbers, excitation level by excitation level}. 

To specify the procedure of switching from the first labeling scheme to the second, we define the \textbf{partition $(\lambda_\vc{t})$ associated with $\vc{t}$}  as follows. Consulting $\vc{t}$ and for  each integer $f=0,1,\ldots,\textsc{t}$, count  
\be
\lambda_f'=(\text{number of modes with $f$ excitations}).
\ee
 The resulting  sequence $(\lambda_0',\lambda_1',\ldots,\lambda_\textsc{t}')$ sums to $N$, the total number of modes. By sorting entries in this sequence in  decreasing order and dropping all $0$ entries, we obtain the partition $(\lambda_\vc{t})$ associated with $\vc{t}$. Excitation numbers $\vc{t},\vc{t}'$ in the same equivalence class always have the same associated partition.
\begin{quote}
\emph{Example:} for $N=4$ and excitation numbers $\vc{t}=(1,3,0,1)$, one obtains the sequence $(1,2,0,1,0,0)$ and thus the associated partition 
$(\lambda_\vc{t})=(2,1,1)$.
\end{quote}

The partition $(\lambda_\vc{t})=(\lambda_1,\lambda_2,\ldots,\lambda_F)$ is represented by a Young diagram: an array of squares where row $f$ has $\lambda_f$ squares. From the Young diagram $(\lambda_\vc{t})$ we obtain the \textbf{Young tableau $\st$ associated with $\vc{t}$} by filling the boxes  with the mode indices from $1$ to $N$ in such a way that mode indices with the same excitation number appear in the same row.
\begin{quote}
\emph{Example:} For $\vc{t}=(1,3,0,1)$,   the Young diagram of $(\lambda_\vc{t})=(2,1,1)$ and a Young tableau associated with  $\vc{t}$ are given by
\[
(\lambda_\vc{t})=
\ytableausetup{centertableaux,boxsize=0.9em}
  		\begin{ytableau}
		 \phantom{0} & \phantom{0}    \\
		 \phantom{0} \\
		 \phantom{0}
		\end{ytableau}
\text{ and } \st=%
\ytableausetup{centertableaux,boxsize=0.9em}
  		\begin{ytableau}
		1 &  4   \\
		3 \\
		2
		\end{ytableau}.
\]
\end{quote}

From the sorting function used to order the entries in the partition $(\lambda_\vc{t})$, one can infer which row in the tableau refers to which excitation number. As a result, the Young tableau lists the mode indices corresponding to each excitation number as intended. One additional modification is required to turn it into the desired state label.

For this, note that the transposition (interchange) of two mode indices with the same excitation number leads to a new tableau (consider interchange of the entries $1$ and $4$ in our example above) but not to a new state $\ket{\vc{t}}$. 
To remove this ambiguity, we define two tableaux as row-equivalent, $\st\overset{\mathsf{R}}{\sim}\st'$ $:\Leftrightarrow$ permutations of elements within each row can transform $\st'$ to $\st$. The resulting equivalence class $[\st]$ is called the \textbf{tabloid associated with $\vc{t}$} and serves as the new label for each state.
\begin{quote}
\emph{Example:} Using $\vc{t}=(1,3,0,1)$ as above, the associated tabloid is
\[
[\st]=%
\left\{
\ytableausetup{centertableaux,boxsize=0.9em}
  		\begin{ytableau}
		1 &  4   \\
		3 \\
		2
		\end{ytableau},
\ytableausetup{centertableaux,boxsize=0.9em}
  		\begin{ytableau}
		4 &  1   \\
		3 \\
		2
		\end{ytableau}
\right\}
\]
\end{quote}
It is useful to note that $[\st]$ can also be expressed as $[\st]=\{ \sigma\st \, |\, \sigma\in\mathsf{R}\st \}$ where $\mathsf{R}\st\subset\SN$ is the subset of permutations which only interchange entries in each row of the tableau $\st$. $\mathsf{R}\st$ called the row-stabilizer of $\st$. Below, we will also encounter the column-stabilizer $\mathsf{C}\st$ with the analogous definition referring to columns instead of rows.

With this, we have established a one-to-one map $\ket{\vc{t}}\leftrightarrow \ket{[\st]}$ which achieves the important goal of relating $\mathcal{V}_{[\vc{t}]}$ to a central object  in the representation theory of $\SN$: the permutation module $M^{\lambda_\vc{t}}$ defined by
\be
M^{\lambda_\vc{t}}=\CC\left\{[\st]\,|\,\vc{t}\in[\vc{t}]\right\}.
\ee
Since each permutation module is defined in terms of tabloids, the group action for  $\mathcal{V}_{[\vc{t}]}$ and for $M^{\lambda_\vc{t}}$ is easily verified to be identical, and the two vector spaces are hence isomorphic as $\SN$ representations.

\subsection*{STEP 2: Decomposing $\mathcal{V}_{[\vc{t}]}$ and constructing basis vectors for all irreducible subspaces} 
The great benefit of identifying $\mathcal{V}_{[\vc{t}]}$ as isomorphic to  $M^{\lambda_\vc{t}}$ lies in the availability of mathematical tools for decomposing the permutation modules into their irreducible subspaces (see Sagan  \cite{Sagan}, section 2.10). For the symmetric group $\SN$, each irreducible representation $S^\mu$ is  labeled uniquely by a partition $(\mu)$ of $N$. Consequently, the general decomposition takes the form
\be
M^\lambda \cong \bigoplus_{\mu}m_{\mu\lambda} S^\mu,
\ee
where $m_{\mu\lambda}\in\NN_0$ is the multiplicity of the irreducible subspace $S^\mu$ within $M^\lambda$. For a given $\mathcal{V}_{[\vc{t}]}\simeq M^{\lambda_\vc{t}}$, we wish to obtain the basis vectors for each of the copies (if any) of $S^\mu$ contained in it. The basis vectors are obtained by means of semi-standard tableaux, which we define next.

For the excitation numbers $\vc{t}$, an \textbf{associated semi-standard tableau $\sst$} is constructed from the Young diagram for $(\mu)$ [where $(\mu)$ need not be $(\lambda_\vc{t})$] by filling its squares with the excitation numbers $t_1,t_2,\ldots,t_N$ in such a manner that entries in each row weakly increase ($\le$), and entries in each column strictly increase ($<$).
Along with $\sst$, we consider \textbf{standard tableaux} $\stref$ of the same shape $(\mu)$, which are Young tableaux with entries increasing in each row and column. This way, we can set up a lookup function $\vartheta:\:\{1,2,\ldots,N\}\to \{t_1,t_2,\ldots,t_N\}$ that extracts the position of the integer $n$ in the reference tableau $\stref$ and returns the excitation number entry found in the semi-standard tableau $\sst$ at the corresponding position. To keep notation simple, we usually do not make the dependence of $\vartheta$ on  $\sst$ and $\stref$ explicit.
\begin{quote}
\emph{Example:}  For $\vc{t}=(0,1,0,3)$ and partition $(\mu) =(2,2)$ an associated semi-standard tableau and standard tableau are: 
\[
\sst=%
\ytableausetup{centertableaux,boxsize=0.9em}
  		\begin{ytableau}
		0 & 0    \\
		1 & 3
		\end{ytableau}, \qquad
		\stref=%
\ytableausetup{centertableaux,boxsize=0.9em}
  		\begin{ytableau}
		1 & 3    \\
		2 & 4
		\end{ytableau}.	
\]
The lookup function then yields the results:
\begin{tabular}{l|llll}
$n$ & $1$ & $2$ & $3$ & $4$\\\hline
$\vartheta(n)$ & 0 & 1 & 0 &3
\end{tabular}. 
\end{quote}

With this preparation, one now obtains the basis states spanning the instance(s) of $S^\mu$ within $\mathcal{V}_{[\vc{t}]}$ from
\begin{align}\label{basiss}
&| \sst;\,\stref  \rangle\\
\nonumber
 &\quad= \sum_{\sigma \in \mathsf{C}\stref } \sum_{\tau \in \mathsf{R}{\stref}} 
	\sgn(\sigma) \sigma\tau
	\left[
	{b^{\dagger}_1}^{\vartheta(1)}{b^{\dagger}_2}^{\vartheta(2)}\cdots {b^{\dagger}_N}^{\vartheta(N)}
	\right] | \vec{0} \rangle
\end{align}
where permutations $\sigma,\tau \in \SN$ act on the mode indices, i.e., the subscripts of the $b^\dag_n$ operators as before. (Caveat: as defined above, the basis states are not normalized yet.)
Each semi-standard tableau $\sst$ yields an irreducible subspace 
\begin{equation}
	\mathcal{V}_{\sst} = \Span \left\lbrace | \sst ;  \stref \rangle 
	\mid \text{all standard tableaux } \stref
	\right\rbrace. 
\end{equation}
By considering all possible partitions $(\mu)$ and associated semi-standard tableaux $\sst$, we thus completely decompose $\mathcal{V}_{[\vc{t}]}$ into linearly independent irreducible subspaces. (The set of partitions one needs to consider can be restricted by considering dominance ordering of partitions  \cite{Sagan}.)

\subsection*{STEP 3: Linear dependence of $b_n$ modes and restricted semi-standard tableaux}
In steps 1 and 2 we have ignored the linear dependence of $b_n$  expressed by the constraint $\sum_n b_n= 0$. Once linear dependence is taken into account, the  states  from Eq.~\eqref{basiss} still span, 
\be
	\mathcal{V}_\TotalEx = \text{span} \left\{ |{\sst};\,\stref \rangle \, \bigg| \, \begin{tabular}{l} $\sst$, $\stref$ semi-standard and \\ standard tableaux, $\TotalEx = \sum_n t_n$ \end{tabular} \right\}.
\ee
However, they are linearly dependent in general. Fortunately, removing this linear dependence can be achieved by a harmless modification of our previous procedure. This modification consists of an additional restriction on the set of admissible semi-standard tableaux $\sst$. Specifically, we will show that restricting the semi-standard tableaux to the set 
\be
\mathcal{R}_\TotalEx = \{ \sst \, | \, \sst \text{ has no ``1" in its first row, }\textstyle\sum_n t_n = \TotalEx\},
\ee
removes the linear dependence and 
\be
	\mathcal{B}_\TotalEx=\left\{ |{\sst};\,\stref \rangle \, \big| \, \sst \in \mathcal{R}_\TotalEx,\; \stref \text{ standard tableau} \right\}
\ee
forms a basis for $\mathcal{V}_{\TotalEx}$. Furthermore, each subspace spanned by states with a fixed restricted semi-standard tableau $\sst$ retains its character as an irreducible representation indexed by the partition $(\mu)$.

To prove this assertion, we first show that every state $|{\sst}';\,\stref\rangle$ obtained for a ``forbidden" semi-standard-tableau ${\sst}' \notin \mathcal{R}_{\TotalEx}$ can be  written as a linear combination of states $\ket{\sst;\,\stref}$ from the restricted set, i.e.\  $\sst \in \mathcal{R}_{\TotalEx}$. 
Consider the vectors constructed in Eq.\ \eqref{basiss} and note that the row-stabilizer can always be separated into the stabilizer of only the first row $\mathsf{R}_1$ and the stabilizer of all remaining rows $\mathsf{R}'$:
\begin{align}
\label{row_stabilizer_decompose}
	\sum_{\tau \in \mathsf{R}\stref} \tau
	= \sum_{\tau' \in \mathsf{R}'} \tau' \circ \sum_{\tau \in  \mathsf{R}_1}  \tau.
\end{align}
\begin{quote} 
\emph{Example}: In this and all following examples we consider the ``forbidden'' state vector
\be
\bigg| 
\ytableausetup{centertableaux,boxsize=0.9em}
\begin{ytableau}
0  & 1 & 2 \\
1  & 4
\end{ytableau} \, ;  
\ytableausetup{centertableaux,boxsize=0.9em}
\begin{ytableau}
1  & 3 & 5 \\
2  & 4
\end{ytableau} \bigg\rangle.
\ee
The stabilizer for row 1 consists of  $\mathsf{R}_1 = \{ e , (13), (15), (35), (135), (153) \} $; the stabilizer for the remaining rows is $\mathsf{R}'=\{ e , (24) \}$. 
\end{quote}

Proceeding with the decomposition of the ``forbidden" state vector $|{\sst}';\,\stref\rangle$ in terms of states with restricted semi-standard tableaux, let 
 $M = \lbrace  m_1,m_2,\cdots,m_N \rbrace$ denote the entries of the standard tableau $\stref$ (reading left to right, row by row), and  $M_1 = \lbrace  m_1,m_2,\cdots,m_r \rbrace$ the entries in row 1 only. Similarly, let $(t_1,t_2,\cdots,t_r)$ be the integer excitation numbers in the first row of the semi-standard tableau ${\sst}'$ and (without loss of generality) assume that $t_1=\cdots=t_{q-1}=0$ and $t_{q} = 1$ for $q\le r$.

Next, we introduce the sets $\Lambda_m=\{n_r,n_{r-1},\ldots,n_m\}$ and rewrite the stabilizer of row 1 as
\begin{align}
\label{stabilizer_to_integers}
	&\sum_{	\tau \in \mathsf{R}_{1}}
	\tau \left[
	b_{m_q}^{\dagger \phantom{t_q}}
	b_{m_{q+1}}^{\dagger t_{q+1}} 
	\cdots 
	b_{m_r}^{\dagger t_r} 
	\right]\\
\nonumber
	&
	= (q-1)! \sum_{n_r \in M_1} \!\!
	b_{n_r}^{\dagger t_r} 
	\sum_{n_{r-1} \in M_1 \setminus \Lambda_r } \!\!\!
	b_{n_{r-1}}^{\dagger t_{r-1}}
    \cdots 
	\sum_{n_{q} \in M_1 \setminus \Lambda_{q+1} } \!\!\!\!
	 b_{n_q}^{\dagger}.
\end{align}
By construction, each successive sum over pseudo-mode indices $n_r,n_{r-1},\ldots$ is associated with weakly decreasing excitation numbers $t_r\ge t_{r-1}\ge\cdots$ and the final sum over pseudo-mode indices $n_q$ corresponds to case of a single excitation (entry ``1" in row 1).
\begin{quote} \emph{Example}: Continuing with our previous example the above equality takes the form 
\begin{align*}
&\sum_{\tau \in \mathsf{R}_1 } \tau \left( b_1^{\dagger 0} b_3^{\dagger} b_5^{\dagger 2} \right) 
=\sum_{n_3  \in \{ 1,3,5 \} }\!\!\!  b_{n_3}^{\dagger 2}  \sum_{ n_2 \in \{1,3,5\} \setminus \{ n_3\} }\!\!\!\!\!\! b_{n_2}^{\dagger} \\
&= b_1^{\dagger 2} \! \left( b_3^\dagger + b_5^\dagger \right) \! + b_3^{\dagger 2} \! \left( b_1^\dagger + b_5^\dagger \right) \! +b_5^{\dagger 2} \! \left( b_3^\dagger + b_1^\dagger \right) \! .
\end{align*}
\end{quote}
Next, we use the linear dependence of the pseudo-modes to rewrite the final sum in Eq.~\eqref{stabilizer_to_integers} as
\begin{align}
\label{use_of_linear_dependence}
\sum_{n_{q} \in M_1 \setminus \Lambda_{q+1}  }\!\!\!  b_{n_q}^{\dagger} 
  = - \sum_{ n \in\Lambda_{q+1} }\!\!\! b_{n}^{\dagger} - \sum_{n \in M\setminus M_1 }\!\!\! b_{n}^{\dagger}. 
\end{align}
The transformed expression has two separate sums over $n$: a sum over pseudo-modes that, according to Eq.~\eqref{stabilizer_to_integers}, are already occupied, and a sum over pseudo-modes in rows $2,3,\ldots$ of the standard tableau. The increase of  excitation number produced by Eq.~\eqref{use_of_linear_dependence} hence only affects pseudo-modes that are already occupied. 
\begin{quote} 
\emph{Example}: Again continuing with our previous example we find
\begin{align}
&\sum_{ n_2 \in \{1,3,5\} \setminus \{ n_3\} }\!\! b_{n_2}^{\dagger} 
=-\sum_{ n \in \{n_3\} }\!\!\!  b_{n}^{\dagger}- \sum_{ n \in \{2,4\} }\!\!\!
 b_{n}^{\dagger}.
  \end{align}
\end{quote}

Finally, we inspect the full state vector by using Eqs.~\eqref{row_stabilizer_decompose}, \eqref{stabilizer_to_integers}, and \eqref{use_of_linear_dependence}. 
The resulting terms associated with a single index $n$ [Eq.~\eqref{use_of_linear_dependence}] can be re-expressed as a sum over the complete row stabilizer $\mathsf{R}$ and associated with a tableaux where the entry ``1" in the first row of ${\sst}'$ has been eliminated and another entry $1\le t_n$ of ${\sst}'$ has been increased by one:
in such a way the state with ``forbidden" semi-standard tableau\\[3mm]
{\footnotesize
\centering
$\ytableausetup{centertableaux,boxsize=3em}
\begin{ytableau}
0  & \cdots & 0 & 1 & t_{q+1} & \cdots & t_{r} \\
t_{r+1} & \cdots & t_{r_2} \\
\vdots
\end{ytableau}
$}\\[3mm]
is decomposed into a linear combination of states with tableaux\\[3mm]
{\centering
\footnotesize
\begin{tabular}{c}
$
\ytableausetup{centertableaux,boxsize=3em}
\begin{ytableau}
0  & \cdots & 0 & \bfr{0} & \, \, t_{q \! + \! 1} \! \bfr{+} \! \! \! \bfr{1} & \cdots & t_{r} \\
t_{r \! + \! 1} & \cdots & t_{r_2} \\
\vdots
\end{ytableau}
$ \\
$\vdots$\\
$
\begin{ytableau}
0  & \cdots & 0 & \bfr{0} & t_{q \! + \! 1} & \cdots & \, \, t_{r} \! \bfr{+} \! \! \bfr{1} \\
t_{r \! + \! 1} & \cdots & t_{r_2} \\
\vdots
\end{ytableau}$ \\
$\phantom{\vdots}$ \\
$
\begin{ytableau}
0  & \cdots & 0 & \bfr{0} & t_{q \! + \! 1} & \cdots & t_{r} \\
\, \, t_{r \! + \! 1 } \! \bfr{+} \! \! \! \bfr{1} & \cdots & t_{r_2}\\
\vdots
\end{ytableau}
$ \\
$\vdots$\\
$
\begin{ytableau}
0  & \cdots & 0 & \bfr{ 0 } & t_{q \! + \! 1} & \cdots & t_{r} \\
t_{r \! + \! 1 } & \cdots & \, \, t_{r_2} \! \bfr{+} \! \! \bfr{1}  \\
\vdots
\end{ytableau} 
$\\
$\vdots$ \, ,\\
\end{tabular}}\\[3mm]
where we have used bold red text to emphasize the changes relative to the ``forbidden'' tableau.  
In cases where the above procedure results in a tableau that is not semi-standard, a straightening algorithm can be applied to generate the corresponding semi-standard tableau  \cite{Sagan}, section 2.6. Importantly the straightening algorithm does not change the content of the tableaux and thus, does not change the fact that our procedure expresses the ``forbidden'' tableau in terms of semi-standard tableau with fewer ``1" entries in row 1. 
Using this procedure, repeatedly if necessary, we can decompose any state with ``forbidden" semi-standard tableau as claimed. It is important, of course, that the removal of states occurs at the level of entire subspaces (indexed by the forbidden semi-standard tableaux) while the group action and hence the irreducibility of the remaining subspaces is unharmed.

\begin{quote} \emph{Example}: We  complete our running example by decomposing the ``forbidden" state vector 
\be
\bigg| 
\ytableausetup{centertableaux,boxsize=0.9em}
\begin{ytableau}
0  & 1 & 2 \\
1  & 4 
\end{ytableau} \, ;  
\ytableausetup{centertableaux,boxsize=0.9em}
\begin{ytableau}
1  & 3 & 5 \\
2  & 4
\end{ytableau} \bigg\rangle.
\ee
in the restricted basis. Following our previous steps, the state vector can be expressed as
\begin{align}
&=\sum_{\sigma \in \mathsf{C}\phantom{'}} \sum_{\tau \in \mathsf{R}'} \text{sgn}(\sigma) \sigma \tau \bigg[ b_2^{\dagger } b_4^{\dagger 4 }\\
\nonumber
&\qquad\bigg( -\sum_{n_3 \in M_1} b_{n_3}^{\dagger 3} - \sum_{n_3 \in M_1} b_{n_3}^{\dagger 2} \sum_{n \in \{ 2 , 4 \} } b_n^\dagger  \bigg) \bigg].
\end{align}
Together with a combinatorial factor (here: $1/2$), the sums over $n_3$ and $\tau'\in\mathsf{R}'$ can be recombined into the full row stabilizer:
\begin{align*}
&=-\frac{1}{2} \sum_{\sigma \in \mathsf{C}} \sum_{\tau \in \mathsf{R}} \text{sgn}(\sigma) \sigma \tau \bigg[ b_2^{\dagger } b_4^{\dagger 4 }
\bigg( b_5^{\dagger 3} + b_5^{\dagger 2} \sum_{n \in \{2,4 \} } b_n^{\dagger}  \bigg)\bigg]\\
&=-\frac{1}{2} \sum_{\sigma \in \mathsf{C}} \sum_{\tau \in \mathsf{R}} \text{sgn}(\sigma) \sigma \tau \bigg[
 b_2^{\dagger } b_4^{\dagger 4 } b_5^{\dagger 3} 
+ b_2^{\dagger 2 } b_4^{\dagger 4 } b_5^{\dagger 2}
+ b_2^{\dagger } b_4^{\dagger 5 } b_5^{\dagger 2} \bigg]
\end{align*}
We have thus completed our goal of expressing the original ``forbidden'' state vector as a linear combination of the ``restricted'' state vectors
\begin{align*}\footnotesize
\bigg| \begin{ytableau}
0  & 0 & 3 \\
1 & 4 
\end{ytableau} \, ;  \begin{ytableau}
1  & 3 & 5 \\
2  & 4
\end{ytableau} \bigg\rangle,
\;
\bigg| \begin{ytableau}
0  & 0 & 2 \\
2 & 4 
\end{ytableau} \, ;  \begin{ytableau}
1  & 3 & 5 \\
2  & 4
\end{ytableau} \bigg\rangle, 
\;
 \bigg| \begin{ytableau}
0  & 0 & 2 \\
1 & 5
\end{ytableau} \, ; \begin{ytableau}
1  & 3 & 5 \\
2  & 4
\end{ytableau} \bigg\rangle.
\end{align*} 
\end{quote}

With the decomposition of ``forbidden" semi-standard tableaux in hand, we conclude by showing that the states $|\sst;\,\stref\rangle$ with restricted $\sst \in \mathcal{R}^\mu_\vc{t}$ not only span each $V_\TotalEx$ but are linearly independent. The proof is based on a simple dimensional argument. 
Counting the number of basis elements with $\TotalEx$ excitations, we find 
\be\label{dimd}
\dim \mathcal{V}_\TotalEx=\frac{(N-2+\TotalEx)!}{\TotalEx! (N-2)!}.
\ee
This should be compared with the dimensionality of subspaces constructed with restricted semi-standard tableaux $\sst\in\mathcal{R}_\TotalEx$. For  $\sum_n t_n = \TotalEx \le 5$, the explicit listing of restricted semi-standard tableaux is given in Table~\ref{RestrictedTableux}.  For comparison with Eq.~\eqref{dimd}, note that the dimension $d_\mu$ of the irreducible representation of $\SN$ indexed by partition $(\mu)$ can be obtained by the hook length formula, see Sagan  \cite{Sagan}, section 3.10. We have verified that
\begin{equation}
\sum_{ \sst \in \mathcal{R}_\TotalEx} d_\mu  = \frac{(N-2+\TotalEx)!}{\TotalEx! (N-2)!}
\end{equation}
for all $\TotalEx \le 5$, and leave it as a conjecture that equality continues to hold for all higher $\TotalEx$.
\begin{quote}
\emph{Example:} As an example we consider $\TotalEx=2$. Using the hook-length formula the sum of the dimension of the irreducible subspaces indexed by the semi-standard tableaux 
\begin{equation}
  	\ytableausetup{boxsize=1em}
  		\begin{ytableau}
		0 & \closedots & 0 & 2 
		\end{ytableau} \,\, , \,\,
  	\ytableausetup{boxsize=1em}
  		\begin{ytableau}
		0 & \closedots & 0  \\
		2 \end{ytableau} \,\, , \,\,
  	\ytableausetup{boxsize=1em}
  		\begin{ytableau}
		0 & \closedots & 0 \\
		1 & 1 \end{ytableau}
\end{equation}
is equal to $1$ + $N-1$ + $N(N-3)/2$. Simple arithmetic shows this is equal to $\dim \mathcal{V}_{\textsc{t}=2} = N(N-1)/2$ as expected.
\end{quote}
In summary,  the subspaces indexed by restricted semi-standard tableaux decompose each $\mathcal{V}_\TotalEx$ into its irreducible subspaces. As a final remark, we note that for $\TotalEx>2$ multiplicities of irreducible representations can exceed $1$. (Through $\TotalEx = 5$, the largest multiplicity that occurs is 5, see Table~\ref{RestrictedTableux}.) In such cases, the usual Gram-Schmidt procedure may be employed to generate orthogonal irreducible subspaces.

\section{Calculation of perturbative shifts according to irreducible subspaces}
\label{IrrepShifts}
We next discuss the calculation of first-order shifts of energy levels in each irreducible subspace under the $\SN$-symmetric perturbation $\dhu$. In general, irreducible subspaces for the difference modes are labeled by restricted semi-standard tableaux. (For  $\TotalEx \le 2$, it is sufficient to specify $\TotalEx$ and the partition $(\lambda)$ instead of full-blown semi-standard tableaux. In this appendix, we continue to employ the restricted semi-standard tableau notation.)

To calculate the first order shifts, we choose a unique element from each subspace. This is done by fixing a \textbf{reference standard tableau} $\Theta^\lambda_\text{ref}$, which we choose as the standard Young tableau of shape $(\lambda)$ with entries $1$ through $N$ filled in column by column. Using this reference tableau, we obtain one representative state in each irreducible difference-mode subspace, which we denote by $|\sstl\rangle$  ($\sstl\in\mathcal{R}^\lambda_\vc{t}$). 
\begin{quote}
\emph{Example:} The reference standard tableau for the partition $(\lambda)=(N-2,2)$ is
\begin{equation}
\Theta^\lambda_\text{ref}=
 \ytableausetup{boxsize=1em}
  		\begin{ytableau}
		1 & 3 & 5 & \closedots & N \\
		2 & 4 \end{ytableau}.
\end{equation}
The state acting as the representative for the $\textsc{t}\!=\!2$, $(\lambda)\!=\!(N\!-\!2,2)$ subspace is then given by
\begin{align*}
\bigg| \, \ytableausetup{boxsize=1em}
  		\begin{ytableau}
		0 & 0 & 0 & \closedots & 0 \\
		1 & 1 \end{ytableau} \, \bigg\rangle 
		&\sim \sum_{\sigma\in \mathsf{C}} \sum_{\tau \in \mathsf{R}} \text{sgn}(\sigma) \sigma \tau \left( {b_2^{\dagger}} {b_4^{\dagger}} \right) | \vec{0} \rangle  \\
		&\sim \left( {b_2^{\dagger}} - {b_1^{\dagger}} \right)\left( {b_4^{\dagger}} - {b_3^{\dagger}} \right)| \vec{0} \rangle, 
\end{align*}
where the column and row stabilizers $\mathsf{C}$ and $\mathsf{R}$ are defined with respect to the reference standard tableau. 
With Eq.~\eqref{b_commutation}, we obtain
\begin{equation*}
	\bigg\langle \, \ytableausetup{boxsize=1em}
  		\begin{ytableau}
		0 & 0 & 0 & \closedots & 0 \\
		1 & 1 \end{ytableau} \,\bigg | \dhu^\text{d} \bigg|  \, \ytableausetup{boxsize=1em}
  		\begin{ytableau}
		0 & 0 & 0 & \closedots & 0 \\
		1 & 1 \end{ytableau} \, \bigg\rangle
		= 2 \ecA (1 - 1/N)
\end{equation*}
for the first-order shift of the irreducible subspace due to the effect of anharmonicity.
\end{quote}

\begin{table*}[H]
\caption{Irreducible subspaces for difference modes with total excitation number $\textsc{t}\le5$. In this table the equivalence class $[{\vc{t}}]$ with $\lambda'_f$ pseudo-modes with exactly $f$ excitations is denoted $0^{\lambda'_0} 1^{\lambda'_1} \cdots$ where all entries with $\lambda'_f=0$ are suppressed. All semi-standard tableau of shape $(\mu)$ and content $[{\vc{t}}]$ are listed in the corresponding row and column.}
\begin{ruledtabular}
\begin{tabular}{| c | c || c | c | c | c | c | c | c | c | c |}
\hline
  $\TotalEx$ & \begin{tabular}{c} $ [ {\vc{t}}  ] $ \\ $0^{\lambda'_0} 1^{\lambda'_1} \cdots $ \\ \end{tabular}  & 
  $(\mu) = (N)$& $(N \! - \! 1,1)$& $(N \! - \! 2,2)$& $(N \! - \! 2,1^2)$& $(N \! - \! 3,3)$& 
  $(N \! - \! 3,2,1)$& $(N \! - \! 4,4)$& $(N \! - \! 4,3,1)$&$(N \! - \! 5, 5)$\\
\hline
\hline
  0  & $0^N$ & 
\tiny
  	\ytableausetup{centertableaux,boxsize=1.2em}
  		\begin{ytableau}
		0  & \closedots & 0
		\end{ytableau} & 
	\phantom{
\tiny
  		\begin{ytableau}
		0  \\
		0
		\end{ytableau}} & 
	&
	&
  	&
  	&
  	&
  	&
	\\
\hline
  1 & $0^{N-1} 1$ & 
  	& 
\tiny
  		\begin{ytableau}
		0  & \closedots & 0 \\
		1
		\end{ytableau} &
	\phantom{
\tiny
  		\begin{ytableau}
		0  \\
		0  \\
		0
		\end{ytableau}} &
	&
  	&
  	&
  	&
  	&
	\\ 
\hline
  2 & $0^{N-1}2$ &
\tiny
  		\begin{ytableau}
		0  & \closedots & 0 & 2
		\end{ytableau} &
\tiny
  		\begin{ytableau}
		0  & \closedots & 0 \\
		2
		\end{ytableau} &
	\phantom{
\tiny
  		\begin{ytableau}
		0  \\
		0  \\
		0
		\end{ytableau}} &
  	&
  	&
  	&
  	&
  	&
	\\
\hline 
    & $0^{N-2}1^2$ &
	&
    &
\tiny
  		\begin{ytableau}
		0 & \closedots & 0 \\
		1 & 1
		\end{ytableau} &
	\phantom{
\tiny
  		\begin{ytableau}
		0  \\
		0  \\
		0
		\end{ytableau}} &
  	&
  	&
  	&
  	&
	\\ 
\hline
  3 & $0^{N-1}3$ &
\tiny
  		\begin{ytableau}
		0  & \closedots & 0 & 3
		\end{ytableau} &
\tiny
  		\begin{ytableau}
		0  & \closedots & 0 \\
		3
		\end{ytableau} &
	\phantom{    
\tiny
  		\begin{ytableau}
		0  \\
		0  \\
		0
		\end{ytableau}} &
    &
  	&
  	&
  	&
  	&
	\\ 
\hline
    & $0^{N-2}1 2$ &
	&
\tiny
  		\begin{ytableau}
		0  & \closedots & 0 & 2 \\
		1 
		\end{ytableau} &
\tiny
  		\begin{ytableau}
		0 & \closedots & 0 \\
		1 & 2
		\end{ytableau} &
\tiny
  		\begin{ytableau}
		0  & \closedots & 0 \\
		1 \\
		2
		\end{ytableau} &
	\phantom{    
\tiny
  		\begin{ytableau}
		0 \\
		0 \\
		0 \\
		0
		\end{ytableau}} &
  	&
  	&
  	&
	\\ 
\hline
    & $0^{N-3}1^3$ &
	&
	&
	&
	&
\tiny
  		\begin{ytableau}
		0  & 0 & \closedots & 0 \\
		1 & 1 & 1
		\end{ytableau} &
	\phantom{    
\tiny
  		\begin{ytableau}
		0 \\
		0 \\
		0 
		\end{ytableau} } &
  	&
  	&
	\\ 
\hline
  4 & $0^{N-1}4$ &
\tiny
  		\begin{ytableau}
		0  & \closedots & 0 & 4
		\end{ytableau} &
\tiny
  		\begin{ytableau}
		0  & \closedots & 0 \\
		4
		\end{ytableau} &
	\phantom{    
\tiny
  		\begin{ytableau}
		0  \\
		0  \\
		0
		\end{ytableau}} &
    &
  	&
  	&
  	&
  	&
	\\ 
\hline
    & $0^{N-2}13$ &
	&
\tiny
  		\begin{ytableau}
		0 & \closedots & 0 & 3 \\
		1
		\end{ytableau} & 
\tiny
  		\begin{ytableau}
		0 & \closedots & 0 \\
		1 & 3
		\end{ytableau} &
\tiny
  		\begin{ytableau}
		0  & \closedots & 0 \\
		1 \\
		3
		\end{ytableau} &
	\phantom{    
\tiny
  		\begin{ytableau}
		0 \\
		0 \\
		0 \\
		0
		\end{ytableau}} &
  	&
  	&
  	&
	\\ 
\hline
    & $0^{N-2} 2^2$ &
\tiny
  		\begin{ytableau}
		0 & \closedots & 0 & 2 & 2\\
		\end{ytableau} & 
\tiny
  		\begin{ytableau}
		0 & \closedots & 0 & 2\\
		2 
		\end{ytableau} & 
\tiny
  		\begin{ytableau}
		0 & \closedots & 0 \\
		2 & 2 
		\end{ytableau} &  
  	\phantom{
\tiny
  		\begin{ytableau}
		0 \\
		0 \\
		0
		\end{ytableau}} &
	&
	&
	&
  	&
	\\ 
\hline
    & $0^{N-3} 1^2 2$ &
	&
	&
\tiny
  		\begin{ytableau}
		0 & \closedots & 0 & 2\\
		1 & 1 
		\end{ytableau} &  
  	&
\tiny
  		\begin{ytableau}
		0  & 0 & \closedots & 0 \\
		1 & 1 & 2
		\end{ytableau} &
\tiny
  		\begin{ytableau}
		0 & \closedots & 0 \\
		1 & 1 \\
		2
		\end{ytableau} &
  	\phantom{
\tiny
  		\begin{ytableau}
		0 \\
		0 \\
		0 \\
		0
		\end{ytableau}} & 
  	&
	\\ 
\hline
    & $0^{N-4}1^4$ &
	&
	&
	&
	&
    &
  	&
\tiny
  		\begin{ytableau}
		0 & 0 & 0 &\closedots & 0 \\
		1 & 1 & 1 & 1
		\end{ytableau} &
	\phantom{
\tiny
  		\begin{ytableau}
		0 \\
		0 \\
		0
		\end{ytableau}} &
	\\ 
\hline
  5 & $0^{N-1}5$ &
\tiny
  		\begin{ytableau}
		0  & \closedots & 0 & 5
		\end{ytableau} &
\tiny
  		\begin{ytableau}
		0  & \closedots & 0 \\
		5
		\end{ytableau} &
	\phantom{    
\tiny
  		\begin{ytableau}
		0  \\
		0  \\
		0
		\end{ytableau}} &
    &
  	&
  	&
  	&
  	&
	\\ 
\hline
    & $0^{N-2}14$ &
	&
\tiny
  		\begin{ytableau}
		0 & \closedots & 0 & 4 \\
		1
		\end{ytableau} &  
\tiny
  		\begin{ytableau}
		0 & \closedots & 0 \\
		1 & 4
		\end{ytableau} &
\tiny
  		\begin{ytableau}
		0  & \closedots & 0 \\
		1 \\
		4
		\end{ytableau} &
	\phantom{    
\tiny
  		\begin{ytableau}
		0 \\
		0 \\
		0 \\
		0
		\end{ytableau}} &
  	&
  	&
  	&
	\\ 
\hline
    & $0^{N-2}2 3$ &
\tiny
  		\begin{ytableau}
		0 & \closedots & 0 & 2 & 3 \\
		\end{ytableau} & 
	\begin{tabular}{c}
\tiny
  		\begin{ytableau}
		0 & \closedots & 0 & 3 \\
		2
		\end{ytableau} \\
		\phantom{.} \\
\tiny
  		\begin{ytableau}
		0 & \closedots & 0 & 2 \\
		3
		\end{ytableau}
	\end{tabular} &  
\tiny
  		\begin{ytableau}
		0 & \closedots & 0 \\
		2 & 3
		\end{ytableau} &
\tiny
  		\begin{ytableau}
		0  & \closedots & 0 \\
		2 \\
		3
		\end{ytableau} &
	\phantom{    
	\begin{tabular}{c}
\tiny
  		\begin{ytableau}
		0 \\
		0 \\
		0
		\end{ytableau} \\
\tiny
  		\begin{ytableau}
		0 \\
		0 \\
		0
		\end{ytableau}
	\end{tabular} } &
  	&
  	&
  	&
	\\ 
\hline
    & $0^{N-3} 1^2 3$ &
	&
	&
\tiny
  		\begin{ytableau}
		0 & \closedots & 0 & 3\\
		1 & 1 
		\end{ytableau} &  
  	&
\tiny
  		\begin{ytableau}
		0  & 0 & \closedots & 0 \\
		1 & 1 & 3
		\end{ytableau} &
\tiny
  		\begin{ytableau}
		0 & \closedots & 0 \\
		1 & 1 \\
		3
		\end{ytableau} &
  	\phantom{
\tiny
  		\begin{ytableau}
		0 \\
		0 \\
		0 \\
		0
		\end{ytableau}} & 
  	&
	\\ 
\hline
    & $0^{N-3} 1 2^2 $ &
	&
\tiny
  		\begin{ytableau}
		0 & \closedots & 0 & 2 & 2\\
		1
		\end{ytableau} &
\tiny
  		\begin{ytableau}
		0 & \closedots & 0 & 2\\
		1 & 2 
		\end{ytableau} &  
\tiny
  		\begin{ytableau}
		0 & \closedots & 0 & 2\\
		1 \\
		2 
		\end{ytableau} &
\tiny
  		\begin{ytableau}
		0  & 0 & \closedots & 0 \\
		1 & 2 & 2
		\end{ytableau} &
\tiny
  		\begin{ytableau}
		0 & \closedots & 0 \\
		1 & 2 \\
		2
		\end{ytableau} &
  	\phantom{
\tiny
  		\begin{ytableau}
		0 \\
		0 \\
		0 \\
		0
		\end{ytableau}} & 
  	&
	\\ 
\hline
    & $0^{N-4}1^3 2 $ &
	&
	&
	&
	&
\tiny
  		\begin{ytableau}
		0 & 0 & \closedots & 0 & 2 \\
		1 & 1 & 1
		\end{ytableau} &
  	&
\tiny
  		\begin{ytableau}
		0 & 0 & 0 & \closedots & 0 \\
		1 & 1 & 1 & 2 \\
		\end{ytableau} &
\tiny
  		\begin{ytableau}
		0 & 0 & \closedots & 0 \\
		1 & 1 & 1 \\
		2
		\end{ytableau} &
	\phantom{
\tiny
  		\begin{ytableau}
		0 \\
		0 \\
		0 \\
		0
		\end{ytableau}}
	\\
\hline
    & $0^{N-5} 1^5$ &
	&
	&
	&
	&
	&
  	&
  	&
  	\phantom{
\tiny
  		\begin{ytableau}
		0 \\
		0 \\
		0
		\end{ytableau}} &
\tiny
  		\begin{ytableau}
		0 & 0 & 0 & 0 & \closedots & 0 \\
		1 & 1 & 1 & 1 & 1
		\end{ytableau}
	\\
\hline
\end{tabular}
\end{ruledtabular}
\label{RestrictedTableux}
\end{table*}

\section{Lagrangian for superinductance and difference mode variables}
\label{DiffModeLagrangian}
After transforming to superinductance and difference mode variables, the Lagrangian of Eq.~\eqref{lag0} can be cast into the form
\begin{align}\nonumber
\mathcal{L}_{\text{SFM}} =&\frac{\hbar^2}{16{\ecsi}}
\dot\phi^2
+\frac{\hbar^2}{16\ecA}\sum_\mu \dot{\xi}_\mu^2+{\ejbs}\cos(\phi + \phx)\\
&+\ejA\sum_{m}\cos\bigg[
\phi/N+ \sum_\mu W_{\mu m} \xi_\mu \bigg].
\end{align}
This expression for the Lagrangian has important advantages over its equivalent form expressed in terms of $\tm$. First,  the kinetic energy is now \emph{diagonal}. Second,
 low-energy minima of the potential energy $U$ are in locations where each difference mode variable vanishes, $\xi_\mu=0$. Third, 
fluctuations between minima are dominantly described by the $\phi$ variable.
The ability to simultaneously expand around $\xi_\mu = 0 $ for each local minimum of $U$ is key in the derivation of the superinductance model used previously  \cite{Manucharyan2009,Manucharyan2010}.

\section{Incorporating  capacitances to ground}
\label{GroundCapAppendix}
The $(N+1)$ node variables $\varphi_j$ can be expressed in terms of the $N$ junction variables $\theta_m$ when using the constraint that the total charge $\sum_j n_j$ of all superconducting islands be zero. To see this, we may use $\tau=\varphi_0$ as a reference variable and express every other node variable $\varphi_j$ ($j=1,\dotsc,N$) as
\be 
\varphi_j = \tau -  \frac{ j}{N}\varphi_\text{ext} +  \sum_{m=1}^{j} \theta_m.
\ee
Note that  $\dot{\tau}\Phi_0/2\pi$ represents a uniform voltage shift of all superconducting islands relative to ground, and that $\tau$ is cyclic, i.e.,  the Lagrangian is independent of $\tau$. Hence, its conjugate momentum is conserved: $\partial {\cal L}/\partial \dot \tau=\text{const}$. This constant of motion, in fact, corresponds to the total charge $n_\text{tot}$ since 
\be
\frac{\partial {\cal L}}{\partial \dot \tau} = \sum_j \frac{\partial \varphi_j}{\partial \tau }\frac{\partial {\cal L}}{\partial {\dot{\varphi}_j}} = \sum_j n_j=n_\text{tot}.
\ee
Imposing the constraint $ n_\text{tot} = 0$ thus allows us to eliminate $\dot{\tau}$ from the Lagrangian, and to work with a Lagrangian (strictly speaking, a Routhian)  which only depends on $\tm$ and $\tmd$. 
Using this procedure the contribution to the kinetic energy due to ground capacitances $\frac{1}{2} (\Phi_0/2\pi)^2 \sum_{i=0}^N C_{\text{g} i} \dot{\varphi}_i^2$ in Eq.~\eqref{lag0} takes the form   $\frac{1}{2}  \sum_{mn} {\cal G}_{mn} \dot{\theta}_m \dot{\theta}_n$ where
\begin{align}
\label{tground}
{\cal G}_{mn} = &\frac{\left( \Phi_0 / 2\pi \right)^2}{\sum_{i=0}^{N} C_{\text{g} i}} \sum_{i=0}^{\text{min}\{m,n\}-1} \! \! \! \sum_{j=\text{max}\{m,n\} }^N \! C_{\text{g}i}C_{\text{g}j}.
\end{align}
Then, assuming the two large superconducting islands surrounding the black sheep junction have ground capacitance $C_\text{g}^\text{b} = e^2/2 E_\text{g}^\text{b}$ while the ground capacitance of the remaining small array islands are $C_\text{g}^\text{a} = e^2 / 2 E_\text{g}^\text{a}$, and using the variables $\xi_\mu$ defined in Eq.~\eqref{difference_def} the kinetic energy terms of the Lagrangian [Eq.~\eqref{lag0}] takes the form
\be 
	\frac{\hbar^2}{16} \sum_{\mu,\nu=0}^{N-1} \left( M_{\mu \nu} + G_{\mu \nu} \right) \dot{\xi}_\mu \dot{\xi}_\nu.
\ee
Here for compactness we use the shorthand $\xi_0 = \phi$. The symmetric $N\!\times\!N$-matrices $M$ and $G$ have the following form:
\be 
M_{00}=1/\ecbs + 1/(N \ecA), \quad M_{0\mu}=0, \quad M_{\mu\nu}=\delta_{\mu\nu}	/\ecA
\ee
and 
\begin{align}
    G_{00} =& \frac{1}{2 \ecgb} + \frac{ (N-1)(N-2)}{12 N \ecga},\\  \nonumber
   	G_{0 \mu}=& - \frac{\text{o}_{\mu+1} \text{c}(\mu)}{2 \ecga \sqrt{2N}\, \text{s}(\mu)^2} \\ \nonumber
    G_{\mu \nu}
    =&
    \frac{\delta_{\mu \nu}}{4 \ecga\,  \text{s}(\mu)^2 } \\
    \nonumber
    &-
    \frac{ \ecgb \, \text{o}_\mu \, \text{o}_\nu \, \text{c}(\mu) \, \text{c}(\nu)}
    {2 N {\ecga}^2 [ 2+ (N-1)\ecgb/\ecga] \text{s}(\mu)^2 \text{s}(\nu)^2},
\end{align}
where $ \text{c}(\mu)$ and $\text{s}(\mu)$ are shorthand for $\cos(\pi \mu /2 N)$ and $\sin(\pi \mu /2 N)$, respectively. Furthermore, the coefficient $\text{o}_\mu$ is $1$ whenever the index $\mu$ is an odd integer, and zero otherwise.    

Performing the Legendre transform, the perturbation from ground capacitances takes the form
\begin{equation}
    \dhc = - 4 \sum_{\mu , \nu = 0}^{N-1} \left[(M+G)^{-1}-M^{-1}\right]_{\mu \nu} \partial_{\xi_\mu}\partial_{ \xi_\nu}.
\end{equation}
For small ground capacitances, the entries of $G$ are small compared to those of $M$ and we approximate $(M+G)^{-1}\simeq M^{-1}-M^{-1}GM^{-1}$, which yields equation \eqref{H_ground_capacitance} in the main text.

\section{Limits of the perturbative approach}
\label{PerturbLimits}
Corrections in the main text are treated perturbatively and we briefly comment on necessary conditions for this approach to be valid. First, we remark that the energy scales $\delta\nspc E_{\text{J}}^\text{sd}$ and $\delta\nspc E_{\text{C}}^\text{sd}$ of Eqs.~\eqref{hcsd} and \eqref{hjsd} must remain sufficiently small relative to the typical energy scales of the superinductance spectrum. Secondly, when the magnitude of $\langle \vec{0} | \dhu^\text{d} | 2 (N) \rangle $ or $\langle \vec{0} |  \dhc^\text{d} (a_1^\dagger)^2 | \vec{0} \rangle $ becomes of order $2 \Omega$, the ground state of the difference modes will requires non-perturbative corrections. To prevent this, the respective inequalities
\begin{eqnarray}
    \sqrt{N/2} \ll 16 / \Delta_\xi^{2},\qquad 
	 (N \Delta_\xi / 2 \pi )^2  \ll	\ecga / \ecA
\end{eqnarray}
must hold. Thus to connect with the $N\rightarrow \infty$ limit (see, e.g., Ref. \cite{Koch2009}) a different approximation scheme to model the low energy spectrum of fluxonium may become necessary.
However, in the case of the fluxonium samples previously studied in Refs. \cite{Manucharyan2009,Manucharyan2010}, the range of validity of the perturbative approach is well satisfied.

\section{Parameter values used in numerical calculations}
\label{ParamValues}
The specific parameters used in all calculations in the main text are $N\!=\!43$, $\ecA\!=\!1.0$, $\ejA\!=\!26.4$, $\ecbs\!=\!3.7$, $\ejbs\!=\!8.9$, $\delta E_{\text{J}}\!=\!0.17 $, $\ecgb\!=\!5$ and $\ecga\!=\!1750$. Using Eqs.~\eqref{renormalized_parameters} these parameters yield $\ecsi \!=\! 2.5$ and $E_L \!=\! 0.53$;  all energies in units of $h\,\text{GHz}$.

%

\end{document}